\documentclass[onecolumn]{aastex631}
\hypersetup{linkcolor=red,citecolor=green,filecolor=cyan,urlcolor=magenta}
\begin{document}

\title{On the parameters of the spherically symmetric parametrized Rezzolla-Zhidenko spacetime through solar system tests, orbit of S2 star about Sgr A$^\star$, and quasiperiodic oscillations}

%% \author[xxxx-xxxx-xxxx-xxxx]{Author Name}

\correspondingauthor{Qiang Wu}
\email{wuq@zjut.edu.cn}

\author{Sanjar Shaymatov}
\affiliation{Institute for Theoretical Physics and Cosmology, Zhejiang University of Technology, Hangzhou 310023, China}
\affiliation{Central Asian University, Milliy Bog Street 264, Tashkent 111221, Uzbekistan}
\affiliation{Institute of Fundamental and Applied Research, National Research University TIIAME, Kori Niyoziy 39, Tashkent 100000, Uzbekistan}
\affiliation{National University of Uzbekistan, Tashkent 100174, Uzbekistan}

\author{Bobomurat Ahmedov}
\affiliation{Institute of Fundamental and Applied Research, National Research University TIIAME, Kori Niyoziy 39, Tashkent 100000, Uzbekistan}
\affiliation{National University of Uzbekistan, Tashkent 100174, Uzbekistan}
\affiliation{Ulugh Beg Astronomical Institute, Astronomy St. 33, Tashkent 100052, Uzbekistan}

%\collaboration{20}{(AAS Journals Data Editors)}
\author{Mariafelicia De Laurentis}
\affiliation{Dipartimento di Fisica, Universitá di Napoli “Federico II”, Complesso Universitario di Monte Sant’Angelo, Edificio G, Via Cinthia, I-80126 Napoli, Italy}
\affiliation{INFN Sezione di Napoli, Complesso Universitario di Monte Sant’Angelo, Edificio G, Via Cinthia, I-80126 Napoli, Italy}

\author{Mubasher Jamil}
\affiliation{School of Natural Sciences, National University of Sciences and Technology, Islamabad 44000, Pakistan}

\author{Qiang Wu}
\affiliation{Institute for Theoretical Physics and Cosmology, Zhejiang University of Technology, Hangzhou 310023, China}
\affiliation{United Center for Gravitational Wave Physics (UCGWP), Zhejiang University of Technology, Hangzhou, 310023, China}

\author{Anzhong Wang}
\affiliation{GCAP-CASPER, Physics Department, Baylor University, Waco, TX 76798-7316, USA}

\author{Mustapha Azreg-A\"{\i}nou}
\affiliation{Engineering Faculty, Ba\c{s}kent University, Ba\u{g}l{\i}ca Campus, 06790-Ankara, Turkey}

\begin{abstract}
	
In this paper, we find the higher order expansion parameters $\alpha$ and $\lambda$ of spherically symmetric parametrized Rezzolla--Zhidenko (PRZ) spacetime by using its functions of the radial coordinate. We subject the parameters of this spacetime to classical tests including weak gravitational field effects in Solar System, observations of the S2 star located in the star cluster close to the Sgr A$^{\star}$ and of the frequencies of selected microquasars. Based on this spherically symmetric spacetime we perform the analytic calculations for Solar System effects like perihelion shift, light deflection, and gravitational time delay so as to determine limits on the parameters by using observational data. We restrict our attention to the limits on the two higher order expansion parameters $\alpha$ and $\lambda$ that survive at the horizon or near the horizon of spherically symmetric metrics. The properties of these two small parameters expansion in PRZ parametrization are discussed. We further apply the Monte Carlo Markov Chain (MCMC) simulations to analyze and obtain the limits on the expansion parameters by using observations of phenomena of the S2 star. Finally, we consider the epicyclic motions and derive analytic expressions of the epicyclic frequencies. Applying these expressions to the quasiperiodic oscillations (QPOs) of selected microquasars allows us to set further limits on the parameters of the PRZ spacetime. Our results demonstrate that the higher order expansion parameters can be given in the range $\alpha\, ,\lambda=(-0.09\, , 0.09)$ and of order $\sim 10^{-2}$ as a consequence of three various tests and observations.  
\end{abstract}

\section{\label{sec:intro}Introduction}
General Relativity (GR), formulated by Albert Einstein in 1915, introduced a new concept of space and time, by showing that massive objects cause a distortion in space-time which is felt as gravity~\citep{Einstein1916}. 
In this way, Einstein's theory predicts, for example, that light travels in curved paths near massive objects, and one consequence is the observation of the Einstein Cross, four different images of a distant galaxy which lies behind a nearer massive object, and whose light is distorted by it. Other well known effects of GR are the observed gradual change in Mercury's orbit due to space-time curvature around the “massive" Sun, or the gravitational redshift, the displacement to the red of lines in the spectrum of the Sun due to its gravitational field.

So far all gravitational field effects in the Solar System (SS) in the weak field approximation and in binary
systems are well described by means of GR \citep[]{Will93}-\citep[]{Will01}.
\textcolor{black}{V. Kagramanova \textit{ et.al.}} \citep{Kagramanova06}
studied SS effects in the Schwarzschild-de Sitter
space-time. Based on this spacetime they calculated SS effects like gravitational redshift, light deflection,
gravitational time delay, perihelion shift, geodetic or de-Sitter precession, as well as the influence of cosmological parameter $\Lambda$ on a Doppler measurements, used to determine the velocity of the Pioneer $10$ and $11$ spacecrafts. Later on, Grumiller constructed an effective model for gravity of a central object at large scales~\citep{Grumiller10}. To leading order in the large radius expansion, he found a cosmological constant, a Rindler acceleration, a term that sets the physical scales and subleading terms. In the last decades, there have been performed hundreds of experiments and observations to check Einstein’s theory. Some of these are entirely new tests, probing aspects of gravity that Einstein himself had never conceived of. One of the first exiting validation of GR was the observation of the binary system PSR B1913 + 16 were was observed that the two stars' obits were shrinking \citep{Hulse74,Hulse75}. This shrinkage is caused by the loss of orbital energy due to gravitational radiation, which is a travelling ripple in spacetime that is predicted by Einstein's GR Theory but never previously verified. 
Very recently we have witnessed numerous gravitational wave observations by the LIGO-Virgo instruments \citep{Abbott16a,Abbott16b}, the study of stars orbiting the supermassive black hole at the center of the Milky Way \citep{Ghez:1998ph},  and the stunning image of the black hole “shadow” in the galaxy M87 \citep{Akiyama19L1,Akiyama19L6}.
In this vast and diverse array of measurements, we have not found a single deviation from the predictions of GR. The string of successes of GR experimentally and observationaly is rather astounding. After more than 100 years, it seems Einstein is still right.

Will this perfect record hold up? We do know, for example, that the expansion of the Universe is speeding up, not slowing down, as recent observations predict~\citep{Di_Valentino_2021}. 
Will this require a radical new theory of gravity, or can we make do with a minimal tweak of GR? As we make better observations of black holes, neutron stars and gravitational waves, will the theory still pass the test? 

Both dark matter and dark energy in the observable Universe has given rise to new alternative theories of gravity since the Einstein's classical GR is not sufficient in explaining those observations \citep{Peebles03,Spergel07,Wetterich88,Caldwell09,Kiselev03,Rubin80,Persic96,Akiyama19L1,Akiyama19L6}. 
Such modified theories of gravity have provided very potent tests in probing new exact solutions for gravitational objects~\citep{Peebles03,Kiselev03,Hellerman01}. 
Thus, it turns out that it is very crucial to parametrize solutions of gravitational field equations. In this framework, a new parametric framework able to mimic space-times geometry of generic static spherically symmetric black holes was proposed in ~\citep{Rezzolla14}. The peculiarity of this approach is to use an expansion of continuous fractions by compactifying  the radial coordinate, which allows a fast convergence. Recently, the  Rezzolla--Zhidenko (PRZ) spacetime has been studied as a model of black hole undergoing spherical accretion of matter and dust \citep{Yang:2020bpj} and as a model of the Galactic center S-stars and distinct pulsars through test particle dynamics~\citep{DeLaurentis18}. Also the PRZ spacetime can be also regarded as a model of the quasiperiodic oscillations observed in microquasars with the low mass X-ray binary systems consisting of either black hole or neutron star and of X-ray data from compact objects~\citep{Bambi12a,Bambi16b,Tripathi19}. These models would play an important role in testing remarkable aspects of PRZ spacetime and in constraining the magnitude of its expansion parameters. Generally, the observed QPOs in the galactic microquasars are determined by the ratio 3:2~\citep{Kluzniak01}, referring to as either high-frequency
(HF) or low-frequency (LF) QPOs with the X-ray power. The HF QPOs are usually referred to as twin peak HF QPOs and provide valuable information on infalling matter in the environment surrounding the compact object. Note that the later ones can get changed in frequency as compared to the HF QPOs which does not tend to drift its frequency~\citep{Tasheva18}. Recent astrophysical observations suggest that LF and HF QPOs arise in the separate parts of accretion disk \citep{Germana18qpo,Tarnopolski:2021ula,Dokuchaev:2015ghx,Kolos15qpo,Aliev12qpo,Stuchlik07qpo,Titarchuk05qpo,Rayimbaev-Shaymatov21a,Azreg-Ainou20qpo,Jusufi21PRD,Ghasemi-Nodehi20qpo,Rayimbaev22qpo}, yet both would be created together in some X-ray binary systems. There are also different models that explain the appearance of HF QPOs in the accretion discs~\citep{Stella99-qpo,Rezzolla_qpo_03a,Torok05A&A,Tursunov20ApJ,Panis19,Shaymatov20egb}. 

In order to  test and compare properties of extended theories of gravity one may use a parameterization that would be able to “mimic" various theories of gravity through expansion of the metric functions in infinitely small dimensionless parameters. First astrophysically interesting perturbative approach based on a perturbations on $M/r$ as deviations of the Kerr metric were developed in \citep{Johannsen11,Glampedakis17}. New parameterization as expansion in small dimensionless distance from event horizon 
was developed for spherically symmetric PRZ metric~\citep{Rezzolla14} and later extended for axially symmetric metric by Konoplya-Rezzolla-Zhidenko (KRZ)
\citep{Konoplya16}.

In this paper, the main idea is to constrain the first two parameters of expansion of PRZ-parameterization with the help of the classical SS tests, the data of the S2 star orbiting the Sgr A$^{\star}$, and the observed frequencies of the QPOs for the selected microquasars. Comparing the theoretical results with observations in the SS, the S2 star, and astrophysical quasiperiodic oscillations observed in GRO J1655-40 and XTE J1550-564 microquasars, we can provide the magnitude of such parameter constraints. Our investigation gives rise to the fact that the constraints provide much information to reveal the properties of such spacetime near the horizon and they can be regarded as a powerful tool in the direction of mimicking the spherically symmetric and slowly-rotating black holes. 

The outline of the paper is as follows. In the
Sec.~\ref{Rezzolla-Zhidenko} we study geodesics of time-like test
particles in the background determined by the spherically
symmetric  PRZ
spacetime~\citep{Rezzolla14}. In Sec.~\ref{perihelionshift}, we study
and derive the expression for the perihelion shift with aim to
obtain the constraints on the first two expansion parameters by comparing
observational data. Secs.~\ref{light bending} and ~\ref{time delay} are devoted to discussing light bending and gravitational time
delay. In Sec.~\ref{S2-star} we discuss constraints on the expansion parameters of spherically symmetric  PRZ spacetime via the S2 star orbit data. We further study the epicyclic motions and aim to constrain on the expansion parameters by applying the
QPOs observed in microquasars in Sec.~\ref{QPOs}. The concluding remarks are given in Sec.~\ref{conclusion}. 

We use in this paper a system of units in which $G_{N}=\hbar=c=1$ (however, for those expressions with an astrophysical application we have written the speed of light and Newtonian gravitational constant explicitly), and is a space-time signature
$(-,+,+,+)$.

\section{\label{Rezzolla-Zhidenko}
	Spherically symmetric  PRZ spacetime}

The line element describing any spherically symmetric spacetime in the Schwarzschild coordinates $(t,r,\theta,\phi)$ is given by \citep{Rezzolla14}
\begin{equation}
	\label{metric}
	ds^2=-g_{tt}(r)dt^2+g_{rr}(r)dr^2+g_{\theta\theta}(r)
	d\Omega^2=-N^2(r)dt^2+\frac{B^2(r)}{N^2(r)}dr^2+r^2
		d\Omega^2\, ,
\end{equation}
where $d\Omega^2 \equiv d\theta^2+\sin^2\theta d\phi^2$, and $N$ and $B$ are functions of the radial coordinate $r$ only with
\begin{eqnarray}
	\label{eq:bh_hor} N(r_0)=0\, .
\end{eqnarray}

Note here that we omit any cosmological effect for the
the line element Eq.~(\ref{metric}) to be an asymptotically flat spacetime from its asymptotic properties. Following PRZ ~\citep{Rezzolla14} we introduce the following dimensionless variable for radial coordinate 
\begin{equation}
	x \equiv 1-\frac{r_0}{r}\, ,
\end{equation}
where $x=0$ and $x=1$ respectively correspond to the location of the event horizon $r=r_0$ and spatial infinity $r=\infty$. Taking into account the above dimensionless variable we can rewrite the function $N$ as follows:
\begin{equation}
	\label{N2} N^2=x A(x)\, ,
\end{equation}
with 
$A(x)>0$\, {for} $0\leq x\leq1$. 
Let us then write the functions $A$ and $B$ in terms of three additional parameters, $\epsilon$, $a_0$, and $b_0$, i.e.,  
\begin{eqnarray}
	\label{asympfix_1}
	A(x)&=&1-\epsilon (1-x)+(a_0-\epsilon)(1-x)^2+{\tilde A}(x)(1-x)^3,\nonumber\\
	\label{asympfix_2} B(x)&=&1+b_0(1-x)+{\tilde B}(x)(1-x)^2\, ,
\end{eqnarray}
with the functions ${\tilde A}(x)$ and ${\tilde B}(x)$ %
%\begin{subequations}

\begin{eqnarray}
	\label{contfrac_1} {\tilde A}(x)=\frac{a_1}{\displaystyle
		1+\frac{\displaystyle
			a_2x}{\displaystyle 1+\frac{\displaystyle a_3x}{\displaystyle
				1+\ldots}}}\, ,\\
	\nonumber \\
	\label{contfrac_2} {\tilde B}(x)=\frac{b_1}{\displaystyle
		1+\frac{\displaystyle
			b_2x}{\displaystyle 1+\frac{\displaystyle b_3x}{\displaystyle
				1+\ldots}}}\, ,
\end{eqnarray}
%\end{subequations}
%
which are proposed to introduce the spacetime at the horizon, $x \simeq 0$ and at the spatial infinity, $x
\simeq 1$. Note that $a_1, a_2, a_3\ldots$ and $b_1, b_2, b_3\ldots$ are the dimensionless constants in the above equations. It is possible for these constants to be constrained on the basis of astronomical observations near the event horizon. Further we obtain the constraints on the first two expansion parameters by using
observational data. From the properties
of the expansions (\ref{contfrac_1}) and (\ref{contfrac_2}), the first two terms at the horizon can be written as 
\begin{eqnarray}
	{\tilde A}(0)={a_1}\, , \qquad {\tilde B}(0)={b_1}\, .
\end{eqnarray}
Therefore the lowest order terms would have primary importance near the horizon. Here, we introduce new
additional parameters $\alpha$ and $\lambda$ to describe only the lowest order terms up to $a_1$ and $b_1$ near the horizon, so that the functions
${\tilde A}(x)$ and ${\tilde B}(x)$ can be expressed as follows: 
\begin{eqnarray}
	\label{asympfix_11}
	x A(x)&=&1-\left(1+\epsilon\right) (1-x)+a_0 (1-x)^2+(a_1-a_0+\epsilon)(1-x)^3- a_1(1-x)^4,\\
	\label{asympfix_22}
	\frac{B^2(x)}{N^2(x)}&=&1+\left(1+\epsilon+2b_0\right)(1-x)+\left(2b_1+2b_0(1+\epsilon)-a_0\right)(1-x)^2\,.
\end{eqnarray}

\subsection{PPN formalism}

An interesting advantage of the PPN formalism is to constrain many theories of gravity. 
Here, we can specifically add the new parameters along with PPN
asymptotic behaviour by representing $B$ and $N$ as
\begin{eqnarray}\label{nsquare}
	\label{expansion_1} N^2 &=&
	1-\frac{2M}{r}+(\beta-\gamma)\frac{2M^2}{r^2}+\alpha\frac{2M^3}{r^3}+
	{\cal O}\left(r^{-4}\right)\nonumber\\
	&=& 1 -\frac{2M}{r_0}(1-x)+(\beta-\gamma)\frac{2M^2}{r_0^2}(1-x)^2+\alpha\frac{2M^3}{r_0^3}(1-x)^3 +{\cal
		O}\left((1-x)^4\right)\,,\\
	\label{expansion_2} \frac{B^2}{N^2} &=& 1
	+\gamma\frac{2M}{r}+\lambda\frac{2M^2}{r^2}+
	{\cal O}\left(r^{-2}\right)\nonumber\\
	&=& 1+\gamma\frac{2M}{r_0}(1-x)
	+\lambda\frac{2M^2}{r^2_0}(1-x)^2+{\cal O}\left((1-x)^3\right)\,,
\end{eqnarray}
where $M$ is the Arnowitt-Deser-Misner (ADM) mass of the spacetime
parameters $\beta$ and $\gamma$ are the  PPN parameters, which are
observationally constrained to be~\citep{Will06}
\begin{equation}\label{PPN}
	|\beta-1|\lesssim2.3\times10^{-4},
	\qquad|\gamma-1|\lesssim2.3\times10^{-5} \, ,
\end{equation}
while $\alpha$ and $\lambda$ are new dimensionless parameters related to the new
coefficients $a_1$ and $b_1$, respectively.

We have expanded the metric function $N(r)$ to
${\cal
	O}\left((1-x)^4\right)$, but kept $B^2(r)/N^2(r) $ to ${\cal
	O}\left((1-x)^3\right)$ as the highest-order constraint on $N(r)$ and $B^2(r)/N^2(r) $ that allow us to find the
parameters $\alpha$ and $\lambda$. So that, the parameter
$\alpha$ sets constraint on $N(r)$ to third order, while $\lambda$
sets constraint on $B^2(r)/N^2(r)$ to second order in $(1-x)$.  By
comparing the two asymptotic expansions
(\ref{asympfix_11})--(\ref{asympfix_22}) and
(\ref{expansion_1})--(\ref{expansion_2}), and collecting terms at
the same order, we find that
\begin{eqnarray}
	1+\epsilon &=& \frac{2M}{r_0}\,, \\
	a_0 &=& (\beta-\gamma)\frac{2M^2}{r_0^2}\,,\\
	1+\epsilon+2b_0 &=&\gamma\frac{2M}{r_0}\,, \\
	a_1+\epsilon-a_0 &=&\alpha\frac{2M^3}{r^3_0}\,, \\
	2b_1+2b_0(1+\epsilon)-a_0
	&=&\lambda\frac{2M^2}{r^2_0}\, .
\end{eqnarray}
Hence, the introduced dimensionless constant $\epsilon$ is completely fixed by the horizon radius $r_0$ and the ADM mass $M$ as
\begin{equation}
	\epsilon=\frac{2M-r_0}{r_0} = - \left(1 - \frac{2M}{r_0}\right)\, ,
\end{equation}
and thus measures the deviations of $r_0$ from $2M$. On the other
hand, the coefficients $a_0$, $b_0$, $a_1$, and $b_1$  can be seen
as combinations of the PPN parameters and new parameters $\lambda$
and $\alpha$ as
\begin{eqnarray} \label{coefficients}
	a_0&=&\frac{(\beta-\gamma)(1+\epsilon)^2}{2}\,, \nonumber\\
	b_0&=&\frac{(\gamma-1)(1+\epsilon)}{2}\, , \nonumber\\
	a_1&=&\left[\beta-\gamma+\frac{\alpha(1+\epsilon)}{2}\right]\frac{(1+\epsilon)^2}{2}-\epsilon\, ,\nonumber\\
	b_1&=&\left[\frac{\lambda}{2}-2(\gamma-1)+\frac{\beta-\gamma}{2}\right]\frac{(1+\epsilon)^2}{2}\, ,
\end{eqnarray}
alternatively, as
\begin{eqnarray}
	\beta  &=& 1 + \frac{2\left[a_0 + b_0(1+\epsilon)\right]}{(1+\epsilon)^{2}}\, ,\\
	\gamma &=& 1 + \frac{2b_0}{1+\epsilon}\, .
\end{eqnarray}

Hereafter we shall study the motion of test particles in the PRZ spacetime Eq.~(\ref{metric}) so as to determine constraints of
parameters with the help of SS effects. Note that
the PRZ black hole spacetime, being static and spherically symmetric, admits two Killing vectors, i.e., $\xi^{\mu}_{(t)}=(\partial/\partial t)^{\mu}$ and
$\xi^{\mu}_{(\phi)}=(\partial/\partial \phi)^{\mu}$ which are referred to as a stationary and an axisymmetry. Hence, there exist two conserved quantities, i.e., the energy $E$ and angular momentum $l$ of the massive particle. Following to these two Killing vectors one can write the generalized momenta as follows:
\begin{eqnarray}
	\label{Enn} -E= g_{\mu\nu}u^{\mu}\xi^{\nu}_{(t)}=-N^2(r)\dot{t}\, ,\\
	\label{Lnn}
	l=g_{\mu\nu}u^{\mu}\xi^{\nu}_{(\phi)}=\frac{B^2(r)}{N^2(r)}\dot{\phi}\, ,
\end{eqnarray}
where $u^{\mu}=\frac{d x^{\mu}}{d\tau}$ refers to the four-velocity of the massive particle with the coordinate four-vector $x^{\mu}$ and the dot denotes derivative with respect to the proper time $\tau$ of massive particle.  
Solving the above equations would give the equations of motion
for a massive particle. Note that we shall further restrict motion to the equatorial plane, i.e. $\theta=\pi/2$. The equation of motion of the massive particle can then be written as  
\begin{eqnarray}\label{eq:mot}
	\dot{t}=\frac{E}{N^2(r)} \mbox{~~and~~} \dot{\phi}=\frac{N^2(r)}{B^2(r)}l\, ,  
\end{eqnarray}
with the one given in terms of the effective
potential as
\begin{eqnarray}\label{eff}
	\frac{\dot{r}^{2}}{2}+V^{\rm{eff}}=E^2\, ,
\end{eqnarray}
where we defined $E={E}/{m}$ and $l={l}/{m}$ are conserved quantitites. Taking all together the effective potential for time-like massive particles reads as
\begin{eqnarray}\label{eff1}
	V^{\rm{eff}}=-\frac{M}{r}-(2-\beta-\gamma)\frac{M^2}{r^2}+\frac{l^{2}}{2r^{2}}
	\left[1-\gamma\frac{2M}{r}\right.-\left.(\lambda-2\gamma^2)\frac{2M^2}{r^2}\right]-\left(4-4\beta+2\beta\gamma+2\gamma^2-2\lambda-\alpha\right)\frac{M^3}{r^{3}}
	\, .
\end{eqnarray}
For light-like test particles, the effective potential simplifies
to
\begin{eqnarray}\label{eff2}
	V^{\rm{eff}}&=&-(1-\gamma)\frac{M}{r}-(2-\beta-\gamma-\lambda+2\gamma^2)\frac{M^2}{r^2}+\frac{l^{2}}{2r^{2}}\left[1-\gamma\frac{2M}{r}
	-(\lambda-2\gamma^2)\frac{2M^2}{r^2}\right]\nonumber\\&&-\left(4-4\beta+2\beta\gamma+2\gamma^2-2\lambda-\alpha\right)\frac{M^3}{r^{3}}
	\, .
\end{eqnarray}
The first term of expression (\ref{eff1}) is the Newton potential,
whereas the second and third terms correspond to the centrifugal
barrier and relativistic correction relevant to the
parameters of spherically symmetric  PRZ
spacetime~\citep{Rezzolla14}. The effective potentials are essential
for the SS effects in order to find constraints on the parameters from observational data.

\section{Limits on the parameters of the PRZ spacetime via SS tests}

\subsection{\label{perihelionshift}
	Perihelion Shift }

In this subsection, we aim to derive the analytic form of the perihelion shift in the case of small eccentricity and small expansion parameters \citep{Wald84}. Particles will oscillate harmonically around some stable circular orbit at radius $r=r_{+}$ with frequency, which is given by 
\begin{eqnarray}
	\omega_{r}^{2}=\frac{1}{r^2_+}\left\{-\frac{2M}{r_{+}}+\frac{l^{2}}{r_{+}^{2}}
	\left[3-12\gamma\frac{M}{r_+}-20(\lambda-2\gamma^2)\right.\right.
	\left.\left.\frac{M^2}{r^2_+}\right]-(2-\beta-\gamma)\frac{6M^2}{r^2_+}\right.-\left.\left(4-4\beta+2\beta\gamma-2\lambda-\alpha\right)\frac{12M^3}{r^3_+}\right\}\,
	.
\end{eqnarray}
The condition, $dV^{eff}/dr\mid_{r=r_{+}}=0$, allows one to determine the angular momentum, and the angular frequency
$\omega_{\phi}=|l|/r_{+}^{2}$, where $l$ is the conserved angular momentum, then takes the form
\begin{eqnarray}
	\omega_{\phi}^2=
	\frac{r^{-2}_{+}}{1-3\gamma\frac{M}{r_+}-4(\lambda-2\gamma^2)\frac{M^2}{r^2_+}}
	\left[\frac{M}{r_+}+(2-\beta-\gamma)\frac{2M^2}{r^2_+}\right.+\left.(4-4\beta+2\beta\gamma-2\lambda-\alpha)\frac{3M^3}{r^3_+}\right]\,
	.
\end{eqnarray}

The perihelion precession is given by
\begin{eqnarray}
	\omega_{p}=\frac{3M^{3/2}}{r_{+}^{5/2}}\bigg\{1+\frac{2\gamma-\beta-1}{3}
	+\frac{1}{6}\frac{M}{r_+}\Big[6(A_{1}+2\lambda)-A^{2}_2+9A_{2}\gamma-6\gamma^2\Big]+O\left(\frac{1}{r_{+}^2}\right)\bigg\}\, ,
\end{eqnarray}
where $A_{1}=4-4\beta+2\beta\gamma+2\gamma^2+2\lambda-\alpha$ and
$A_{2}=2-\beta-\gamma$. 
\begin{table*}
	\centering
	\begin{tabular}{||c||c|c|c|c|c|c|c|c|c||} \hline\hline
		Planet & Mercury & Venus & Earth & Mars & Jupiter & Saturn &
		Uranus & Neptune & Icarus  \\\hline
		$P$ & $4\cdot 10^{45}$ & $7\cdot 10^{45}$  & $9 \cdot 10^{45}$ & $1.4\cdot 10^{46}$ & $5\cdot 10^{46}$ & $9\cdot 10^{46}$ & $1.8\cdot10^{47}$ & $3\cdot 10^{47}$ & $1.0\cdot10^{46}$ \\
		$e$ & 0.2 & 0.007 & 0.017 & 0.09 & 0.05 & 0.06 & 0.05 & 0.011 & 0.8 \\
		$\triangle\phi=2\pi\omega_p P^{3/2}/\sqrt{M}$ & 43 & 8.6 & 3.8 & 1.3 & 0.06 & 0.014 & 0.002 & 0.0007 & 9.8 \\
		$\delta\omega_p/\omega_p$ & $1.2\cdot 10^{-4}$ & $3\cdot 10^{-2}$
		& $1.1\cdot 10^{-4}$ & $4\cdot 10^{-4}$ & $0.6$ & $3\cdot 10^{2}$
		& $7\cdot 10^3$ & ? & $8\cdot10^{-2}$\\ \hline\hline
	\end{tabular}
	\caption{Numerical values of semimajor axes $P$, eccentricities $e$, perihelion shifts $\triangle\phi$, and the relative perihelion shifts $\delta\omega_p/\omega_p$ are tabulated for all SS planets \citep{Grumiller11}. We do however note that the values of the uncertainties in perihelion shifts of SS planets have been discussed and presented recently in Refs.~\citep{Iorio2019ApJ,Iorio15IJMPD}. }
	\label{tab1}
\end{table*}
Following \citep{Weinberg72}, the perihelion precession in the above equation can be written in terms of finite eccentricity and the semimajor axis as follows:
\begin{eqnarray}\label{p.shift1}
	\omega_{p}=\frac{3M^{3/2}}{(1-e^{2})P^{5/2}}\bigg\{1+\frac{2\gamma-\beta-1}{3}
	+\frac{1}{6}\frac{M}{r_+} \Big[6(A_{1}+2\lambda)
	-A^{2}_2+9A_{2}\gamma-6\gamma^2\Big]+O\left(\frac{1}{r_{+}^2}\right)\bigg\}\, ,
\end{eqnarray}
where $P$ and $e$ respectively refer to the semimajor axis  and eccentricity of the ellipse. After some straightforward calculations, we obtain the perihelion shift as follows
\begin{eqnarray}\label{p.shift2}
	\triangle\phi=\frac{6\pi
		M}{(1-e^{2})P}\bigg\{1+\frac{2\gamma-\beta-1}{3}
	+\frac{1}{6}\frac{M}{r_+} \Big[6(A_{1}+2\lambda) -A^{
		2}_2+9A_{2}\gamma-6\gamma^2\Big]+O\left(\frac{1}{r_{+}^2}\right)\bigg\}\, ,
\end{eqnarray}
where the leading order term in Eq.~(\ref{p.shift2}) must be small in order not to conflict with observational data.
Considering now the observational data \citep{Grumiller11} tabulated in Table \ref{tab1} and using the PPN bound, we try to gauge the value of the residual perihelion shifts of the planets Mercury $\delta\omega_p/\omega_p$ in particular cases. With the solar mass given by $M\approx 9\cdot 10^{37}$ in Planck units the third term of the expression Eq.~(\ref{p.shift2}) will be extremely small 
\begin{eqnarray}\label{bound}
	\frac{M}{r_+}(4\lambda-\alpha)\ll 1 \, , %10^{-3}
\end{eqnarray}
due to the fact that $M/r_+\sim M/P\approx 10^{-8}$. For that the perihelion shift data becomes very weak and does not allow one to get strong constraints on the value of expansion parameters $\lambda$ and $\alpha$ together with the PPN parameters. Hence, the bound from perihelion shift cannot give strong constraints on the above mentioned parameters. Further (\ref{bound}) implies that: 
\begin{eqnarray}\label{newppn1}
	&& \lambda \ll 1 \mbox{~~and~~ } \alpha \ll 1\, .
\end{eqnarray}

\subsection{\label{light bending} Light Bending}

In this subsection, we study and explore the classical light bending. For that we consider the quantity $\dot{r}$ in Eq.~(\ref{eff}) which vanishes at the point where the worldline of photon becomes close to the sun, i.e., $r=r_{0}$. Thus, the effective potential given by Eq.~(\ref{eff2}) is taken to be the energy $E$ at $r_{0}$. Putting this value for $E$ into Eq.~(\ref{eff}) with the photon trajectory one can then have the following expression as
\begin{eqnarray}
	\frac{dr}{d\phi}=\frac{r N(r)}{B(r)}\left[\left(\frac{r
		N(r_0)}{r_{0}N(r)}\right)^2-1\right]^{1/2}\, ,
\end{eqnarray}
where $N(r)$ and $B(r)$ functions are given in Eq.~(\ref{nsquare}). Now it is straightforward to obtain a photon's deflection angle as follows:
\begin{eqnarray}\label{bending}
	\Delta \varphi&=&2\int_{r_{0}}^{\infty}\frac{dr}{\frac{r
			N(r)}{B(r)} \left[\left(\frac{r
			N(r_0)}{r_{0}N(r)}\right)^2-1\right]^{1/2}}-\pi\nonumber\\
	&=&\frac{4M}{r_{0}}\left\{
	\frac{1+\gamma}{2}-\gamma\frac{M^2}{r^{2}_0}+\frac{(3\gamma-\beta)}{4}\frac{M^2}{r^{2}_0}\pi
	+\frac{\lambda}{8}\frac{M^2}{r^{2}_0}\pi\right.-\left.\left[\gamma(\beta-\gamma)+\alpha+\frac{\alpha}{4}\pi\right]\frac{M^3}{2r^{3}_0}+O\left(\frac{M}{r_{0}}\right)^{4}\right\}\nonumber\\
	&=&\frac{4M}{r_{0}}\left(\frac{1+\gamma}{2}\right)+\delta\Delta\varphi \, .
\end{eqnarray}
Here we note that the term, $4M/r_{0}$, refers to the general relativistic effect and the rest of all terms in the brackets of Eq.~(\ref{bending}) must be sufficiently small for not conflicting with observational data. Now we take into account the observational data for the deflection angle \citep{Grumiller11,Shapiro04PRL}, which allow one to constrain the expansion parameters $\lambda$ and $\alpha$ through the bound on the deflection angle
\begin{eqnarray}\label{bounddeflection}
	-1.2\times10^{-3}<\frac{\delta\Delta\varphi}{\Delta\varphi}< 4\times10^{-4}\, . 
\end{eqnarray}
The term $\delta\Delta\varphi$ in the above expression includes the second and third order terms in $M/r_{0}$. Following to \citep{Grumiller11}, $r_0$ can be approximately taken to be two solar radius, i.e., $r_0\approx 8\times 10^{43}$ in Planck units. Plugging this value into Eq.~(\ref{bounddeflection}) one can constrain on the expansion parameter $\alpha$ and $\lambda$. However, it turns out to be very complicated to get the constraints on these parameters. In the fact that $\delta\Delta\varphi$ including higher order terms of $M/r_{0}$ could be equal to very small value, which is of order $\left(M/r_{0}\right)^2\approx 10^{-12}$ and $\left(M/r_{0}\right)^3\approx 10^{-18}$. It does therefore turn out that the deflection angle with the observational data could not provide a strong constraints on the above mentioned parameters. One can however expect that these parameters take small values, i.e., $\lambda \sim \alpha\ll 1$.

\subsection{\label{time delay}%Gravitational time delay
Radar echo delay}

We then turn to the standard measurement of the time delay that stems from the light bending. Here, we consider the time delay of light and evaluate it with together clock effects for a radar signal, which is sent from Earth to some object and is reflected back from target to Earth (see for example \citep{Weinberg72}). For that we shall first consider the following differential equation, which can be written as follows:  
\begin{eqnarray} \label{timedelay1}
	\frac{dr}{dt}=\frac{N^2(r)}{B(r)}\bigg[1-\left(\frac{N(r)r_{0}}{N(r_0)r}\right)\bigg]^{1/2}\, 
	.
\end{eqnarray}
From the above equation, we calculate the integral for the time, for
which the light travels from $r_0$ to $r$. Hence, it will have the form as
\begin{eqnarray} \label{timedelay2}
	t(r,r_{0})&=&\int_{r_0}^{r}\left(1-\frac{r_0^{2}}{r^2}\right)^{-1/2}\left\{1+(1+\gamma)\frac{M}{r}\right.+\frac{M r_0}{r(r+r_0)}+
	(4\gamma-2\beta+\lambda)\frac{M^2}{r^2}\nonumber\\&&+\left[2\lambda-2\gamma(\beta-\gamma)-2\alpha-2(\gamma+1)(\beta-\gamma)\right]\frac{M^3}{r^3}-\left.\alpha\frac{M^3}{r_0r^2}+\alpha\frac{M^3}{r^2(r+r_0)}\right\}dr\,
	,
\end{eqnarray}
and we then find the time delay in general form as follows: 
\begin{eqnarray}\label{timedelay2}
	\Delta t&=&
	2\left[t(r_{E},r_{0})+t(r_{T},r_{0})-\sqrt{r^{2}_{E}-r^{2}_{0}}\right.-\left.\sqrt{r^{2}_{T}-r^{2}_{0}}\right]\nonumber\\&=&
	4M\left\{1+\frac{1+\gamma}{2}\ln\frac{4r_{E}r_{T}}{r_{0}^{2}}\right.+(4\gamma-2\beta+\lambda)\frac{M}{4
		r_0}\pi+\left.\big[\lambda-\gamma(\beta-\gamma)-2\alpha-(\gamma+1)(\beta-\gamma)\big]\frac{M^2}{r^{2}_0}+O\left(\frac{M}{r_{0}}\right)^{3}\right\}\nonumber\\
	&=&4M\left(1+\frac{1+\gamma}{2}\ln\frac{4r_{E}r_{T}}{r_{0}^{2}}\right)+\delta\Delta
	t\ .
\end{eqnarray}
It is worth noting here that for further analysis we consider $r_{0}$ as a distance, which is of the solar radius order. Also, $r_{T}$ and $r_{E}$ respectively refer to the distance of the target and the Earth from the sun. However, we shall for simplicity consider $r_{T}$ and $r_{E}$ as the semimajor axes of the target
and the Earth orbits for further analysis; see Table~\ref{tab1}. We note that in the above equation (\ref{timedelay2}) the first term on right hand side corresponds to the general relativistic PPN result, while the second term corresponds to the correction that comes from the expansion parameters of spherically symmetric  PRZ spacetime. In doing so, we analyse the time delay and its bound in order to obtain both upper and lower limits on the expansion parameters $\lambda$ and $\alpha$ using the observational data \citep{Bertotti03Nat}
\begin{eqnarray}
	-10^{-6}<\frac{\delta\Delta t}{\Delta t}< 2\cdot 10^{-5} \, ,
\end{eqnarray}
which, in turn, allows one to determine the bound for higher order expansion parameters associated with the higher order terms of $M/r_{0}$. However, $M^2/r_{0}^2$ in the third term of Eq.~(\ref{timedelay2}) is of order $\approx 10^{-12}$ in the case of the distance taken to be $r_0\approx7\times10^{43}$ (see for example \citep{Bertotti03Nat}). It does not therefore cause explicit constraints on these higher order expansion parameters. As a consequence of analysis of the time delay due to light bending, one can deduce that these higher order expansion parameters could take small values, i.e., $\lambda\sim\alpha\ll 1$, similarly to what is observed in the previous analysis.

In this section, we have shown that it is not possible to obtain accurate and explicit constraints on the higher order expansion parameters, $\alpha$ and $\lambda$ through the observational data for the solar system tests. This happens because the factor $M/r_{+}$ and $M/r_{0}$ and their higher order terms in the expression of perihelion shift, light deflection, and gravitational time delay are infinitely small and thus have no significant impact on SS tests. Therefore, the observational data for SS tests are not able to have enough capacity to put a strong constrain on the higher order expansion parameters of the  PRZ spacetime. To obtain the best-fit constraints on these expansion parameters we must test and consider observations of phenomena of
the S2 star located in the star cluster close to the Sgr A$^{\star}$. It would therefore be expected that the observational data for the S2 star orbiting around the Sgr A$^{\star}$ have enough capacity to constrain the higher order expansion parameters. This is what we intend to examine in the next section.

\section{Limits on the parameters of the PRZ spacetime via the star S2 orbit data \label{S2-star} }

The galactic center of the Milky Way galaxy provides an excellent physics laboratory for various observational and experimental tests of the black hole models and theories of modified gravity in their most extreme limits. In fact, it was argued that one can use the S cluster stars to set constrains on the BH mass which is $4.07\times10^6 M_\odot$ at a distance approximately 8.35 kiloparsec (or kpc) from us \citep{Gillessen17ApJ}. So far there is no observational evidence if Sgr A$^{\star}$ possesses an astrophysical jet or any accretion disk. Moreover it is less clear if Sgr A$^{\star}$ is accreting gas via Bondi radially infall accretion or through a disk.  However, the indirect evidence of accretion disk about Sgr A$^{\star}$ is relevant to the recent detection of Sgr A$^{\star}$ shadow by Event Horizon Telescope (EHT) collaboration \citep{Akiyama22ApJL}. The shadow is formed by the light emitted from the accretion disk, subsequently deflection by BH and eventually arriving at EHT detectors. Further constraints on the mass, spin and geometry of Sgr A$^{\star}$ BH would be deduced from the observations of its shadow in the near future \citep{Perlick21}. 

In this section,  we focus on the constraints of the expansion parameters by using observations of phenomena of the star S2 located in the star cluster close to Sgr A$^{\star}$. The star S2 orbits the Sgr A$^{\star}$ \citep{Lacroix18,Nucita07,Ghez05,Ghez00}, and thus it plays an important role to study the physical properties of the central black hole, Sgr A$^{\star}$, i.e., its parameters being of the primary astrophysical importance. It has been shown very recently that wormhole models as candidates for Sgr A$^{\star}$ are tested by the motion of the S2 star (see for example \citep{Jusufi22EPJC}). Moreover, the motion of S2 star has been used to constrain different models for the dark matter distribution inside the inner galactic region, such as the dark matter spike model~\citep{Nampalliwar21ApJ}, loop quantum gravity model~\citep{Yan22JCAP}. Also, the S2 star observations have been used to constrain barrow entropy~\citep{Jusufi22Univ}. With this motivation, we also consider the motion of the S2 star to set constraints on parameters of PRZ black hole spacetime, as described by the line element proposed in Ref.~\citep{Rezzolla14}, as a candidate for the Sgr A$^{\star}$ compact object at the centre of the Milky Way galaxy. 

Thus, we shall further consider the motion of S2 star around the  PRZ spacetime geometry and solve the equations of motion numerically (see \citep{Becerra-Vergara20} for details) by adapting the method related to the analysis of the periastron shift of the S2 star possessing orbit parameters. Thus, we limit the values of the expansion parameters of spherically symmetric  PRZ spacetime through phenomena of the S2 star. In Fig.~\ref{fig:orbit} we show the best-fitting orbit of the S2 star as stated by the observational data given in Ref.~\citep{Do19}. Note that the star ``$\star$'' in Fig.~\ref{fig:orbit} indicates the location of Sgr A$^\star$. 
\subsection{Datasets \label{Sec:data}}

\begin{figure}
	\centering
	\includegraphics[width=8.1cm]{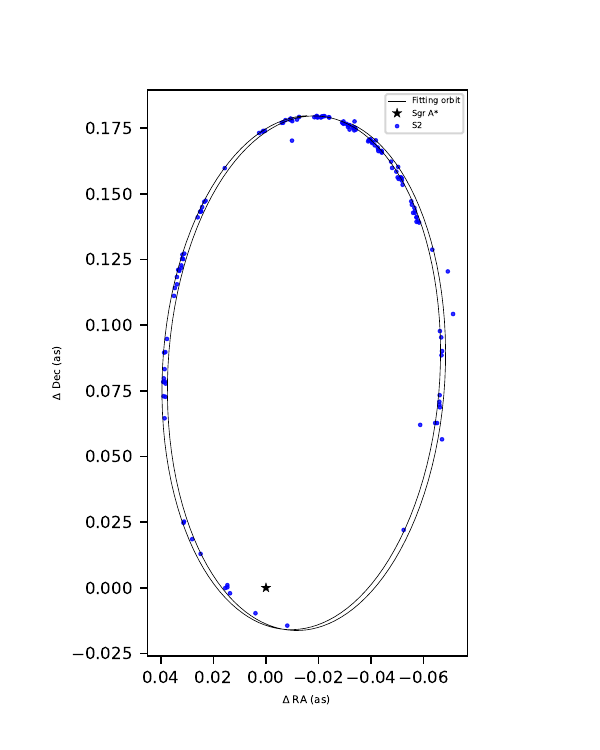}
	\caption{Observed orbit of S2 star around Sgr A$^{\star}$. Note that we obtained the S2 star orbit around Sgr A$^{\star}$ as a model fitting with  PRZ spacetime geometry by applying the observational data. 
	}  \label{fig:orbit}
\end{figure}

\begin{figure}
	\centering
	\includegraphics[width=10.1cm]{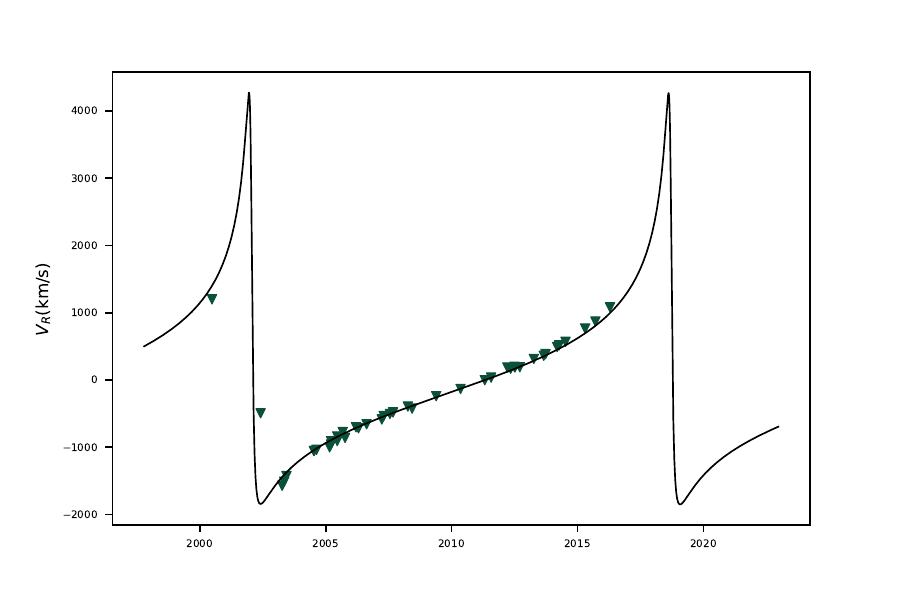}
	\caption{The dataset of radial velocities of S2 star used in our analysis and the fitting curve by PRZ spacetime.}  \label{fig:velocity}
\end{figure}

We further apply the astrometric and spectroscopic data for the S2 star. We note that these data are publicly available and have been collected. There are three various parts for these data, i.e., astrometric positions, radial velocities and the pericenter precession which we further use for our purpose. The details of these three parts can be summarized as follows: 1. {\em Astrometric positions}: 145 astrometric positions for S2 star starting from 1992.224 to 2016.53 as reported in \citep{Gillessen17ApJ} are used in our analysis. It is worth noting that, before 2002 all data had been obtained/collected from the ESO New Technology Telescope (NTT), while the rest part of the data has been collected since 2002 from the Very Large Telescope (VLT). In Fig.~\ref{fig:orbit}, we demonstrate these mentioned data. We show the data from NTT and VLT by blue and red points, respectively, as seen in Fig.~\ref{fig:orbit}. 
2. {\em Radial velocities}: the data for 44 radial velocities can be used from 2000.487 to 2016.519 (see for example \citep{Gillessen17ApJ}). Similarly, the data for radial velocities were collected by various observing sources, i.e., the data from NIRC2 before 2003 and from the INtegral Field Observations in the Near Infrared (SINFONI) after 2003.  
We show these data in Fig.~\ref{fig:velocity} for the radial velocity $V_{R}$ as a function of Epoch year.   
3. {\em Orbital} precession of S2 star: the orbital precession of S2 star has been observed and measured by the GRAVITY Collaboration (see for example \citep{GRAVITY2}). 
\begin{eqnarray}
	\Delta \phi _{per \;orbit}=1.10 \pm 0.19 \label{OP_12}\, . 
\end{eqnarray}
It is well-known fact that the orbital precession comes into play to be important phenomenon as a remarkable prediction of GR; see Fig.~\ref{fig:orbit}. For our purpose, we apply its measurement in the analysis of the Monte Carlo Markov Chain simulations.

%\st{It is worth noting that the epoch is expressed in Julian year.}

\subsection{Modeling the orbit with relativistic effects}

We consider the equations of motion for massive particles and apply them to explore the motion of the S2 star orbit with relativistic effects. For that one needs to integrate Eqs.~(\ref{eq:mot}) and (\ref{eff}) numerically by imposing the initial conditions, including coordinates $\{t(\lambda_0), r(\lambda_0), \phi(\lambda_0)\}$ and their first derivatives $\{\dot t(\lambda_0), \dot r(\lambda_0), \dot \phi(\lambda_0)\}$. This would play a key role in deriving the S2 star positions at the orbital plane, yet its astrometric positions as stated above can be determined at the celestial plane. 
The point to be noted here is that one can approximate the S2 star motion in the above mentioned two planes by using elliptical orbit. It is then possible to make a comparison between the theoretical and the astrometric positions if and only if one considers all observational quantities being in the same plane. For that purpose, one can consider a projection of the theoretical positions on the celestial plane with the help of the following coordinate transformations 
\begin{eqnarray}
	X &=& x \mathcal{B}+y \mathcal{G}\label{Xobs}\, ,\\
	Y&=&x\mathcal{A}+y\mathcal{F}\label{Yobs}\, ,\\
	Z&=&x\mathcal{C}+y\mathcal{H} \label{Zobs}\, .
\end{eqnarray}
Here, we note that $(X, Y, Z)$ and $(x, y, z)$ respectively refer to the coordinates on the celestial and the orbital planes, while $\mathcal{A}, \mathcal{B}, \mathcal{C}, \mathcal{F}, \mathcal{G}, \mathcal{H}$ are the corresponding coefficients that are defined by 
\begin{eqnarray}
	\mathcal{A}&=&\cos \Omega^{\prime} \cos \omega^{\prime} -\sin \Omega^{\prime} \sin \omega^{\prime} \cos i\, . \label{A}\\
	\mathcal{B}&=&\sin \Omega^{\prime} \cos \omega^{\prime} +\cos \Omega^{\prime} \sin\omega^{\prime} \cos i\, , \label{B}\\
	\mathcal{C}&=&\sin \omega^{\prime} \sin i\, ,  \label{C}\\
	\mathcal{F}&=&-\cos\Omega^{\prime} \sin \omega^{\prime} -\sin \Omega^{\prime} \cos \omega^{\prime} \cos i\, , \label{F}\\
	\mathcal{G}&=&-\sin \Omega^{\prime} \sin \omega^{\prime} +\cos \Omega^{\prime} \cos \omega^{\prime} \cos i\, ,  \label{G}\\
	\mathcal{H}&=&\cos \omega^{\prime} \sin i\, , \label{H}
\end{eqnarray}
where $\omega^{\prime}$ and $\Omega^{\prime}$ respectively refer to the perihelion argument and the longitude of ascending node with the orbital inclination $i$ of the S2 star's the elliptical orbit.

We shall further assume that there exists an offset between the gravitational center and the reference frame considered here, and thus we have $x_{0},y_{0},v_{x0},v_{y0}$ for further modeling process (see for example \citep{Do19})
\begin{eqnarray}
	X&=&X(t_{\rm em})+x_{0}+v_{\rm x0}(t_{\rm em})(t_{\rm em}-t_{\rm J2000})\, ,  \label{x0} \\
	Y&=&Y(t_{\rm em})+y_{0}+v_{\rm y0}(t_{\rm em})(t_{\rm em}-t_{\rm J2000})\, .  \label{y0}
\end{eqnarray}
Here we would like to mention that $t_{\rm J2000}$ and $t_{\rm em}$ are used to delineate Julian 2000 year and the epoch of the emitting light, respectively. To compare these theoretical positions with the astrometric data one also needs to take several relativistic effects into account. 

Le us fist take the Romer time delay effect into consideration as it changes the arrival time of the light that comes from the orbiting star located far way or closer to the Earth. We define the Romer time delay as follows:
\begin{eqnarray}
	t_{\rm obs}-t_{\rm em}=\frac{Z(t_{\rm em})}{c}\, ,
\end{eqnarray}
with the epoch $t_{\rm obs}$ in case the light is observed, while $Z$ is derived from Eqs.~(\ref{Zobs}). However, it is very complicated to solve this equation, yet there exists an alternative method to approach and solve this equation (see for details \citep{Do19,GRAVITY1}), and it is given by 
\begin{eqnarray}
	t_{\rm em}^{(i+1)}=t_{\rm obs}-\frac{Z(t_{\rm em}^{(i)})}{c}\, .
\end{eqnarray}
After some iterations, this yields
\begin{eqnarray}
	t_{\rm em} \approx t_{\rm obs}-\frac{Z(t_{\rm obs})}{c}\, .
\end{eqnarray}

Next, we wish to consider the effect of frequency shift $\zeta$ for photon, which has the following relation associated with the S2 star's radial velocity as 
\begin{eqnarray}
	\zeta = \frac{\Delta \nu}{\nu} = \frac{\nu_{\rm em}-\nu_{\rm obs}}{\nu_{\rm obs}}=\frac{V_{\rm R}}{c}\, ,
\end{eqnarray}
with the emitted frequency $\nu_{\rm em}$ and the observed frequency $\nu_{\rm obs}$ of the light, and the S2 star's radial velocity $V_{\rm R}$.  There exist two main relativistic effects that can have a significant impact on the above mentioned frequency shift, i.e., the Doppler shift $\zeta_{\rm D}$ and the gravitational redshift $\zeta_G$ effects. For example, the first one we consider is the Doppler shift $\zeta_{\rm D}$ effect that stems from the relative motion occurring between the observer and the star. Thus, the Doppler shift effect can give rise to a significant impact as that of the S2 star's high orbital velocity 
\begin{eqnarray}
	\zeta_{\rm D}=\frac{\sqrt{1-\frac{v_{\rm em}^{2}}{c^{2}}}}{1- n \cdot  v_{\rm em}}\, .
\end{eqnarray}
Here, we note that $v_{\rm em}$ refers to the velocity measured at at $t_{em}$, while $n \cdot v_{\rm em}$ refers to the the projected velocity onto the light sight (i.e., the radial velocity). Another relativistic effect is the  gravitational redshift effect $\zeta_{\rm G}$ which is caused by pure GR effect, and its frequency shift can be drastically influenced by strong gravitational field 
\begin{eqnarray}
	\zeta_{\rm G}=\frac{1}{\sqrt{-g_{tt}}}\, .
\end{eqnarray}
Since the frequency shift has the relevance with the Doppler and the gravitational field effects it can then be defined by 
\begin{eqnarray}
	\zeta= \zeta_{\rm D} \cdot \zeta_{\rm G}-1\, .
\end{eqnarray}
We further need to describe the corresponding radial velocity $V_{\rm R}$ of the S2 star. For that purpose we consider that any movement of Sgr A$^{\star}$ shifting towards the Sun would lead to the change in the velocity, i.e, there may exist an offset and drift between the reference frame and the gravitational center. To model $V_{\rm R}$ one needs to introduce new parameter $v_{z0}$. One can then be able to define the radial velocity of the S2 star via $v_{z0}$, similarly to what is done previously for the positions given by Eqs.~(\ref{Xobs}-\ref{Zobs}). Hence, it can be written as follows~\citep{Reid07ApJ}:   
\begin{eqnarray}
	V_{\rm R}=c \cdot \zeta + v_{\rm z0}\, .
\end{eqnarray}

\subsection{Analysis of Monte Carlo Markov Chain}

For the further analysis, we apply the MCMC simulations (see for example \citep{emcee}) in order to constrain on the expansion parameters of  spherically symmetric  PRZ spacetime. Let us then introduce the set of parameters  
\begin{eqnarray}
	\{M, R_{0}, a, e, i, \omega^{\prime}, \Omega^{\prime}, t_{\rm apo}, x_{0}, y_{0}, v_{x_{0}}, v_{y_{0}}, v_{z_{0}},\beta,\gamma,\alpha,\lambda\}\label{paras}\, ,
\end{eqnarray}
and explore them to obtained their best fit with the available data associated with the theoretical orbits,  as presented in Sec.~\ref{Sec:data}. Here we note that $M$ and $R_0$ denote the black hole mass and the distance from the black hole to the Earth respectively, while $\{ a, e, i, \omega^{\prime}, \Omega^{\prime}, t_{\rm apo}\}$ represent the set of six orbital elements of the S2 star's elliptical orbit and another set of five parameters $\{x_{0}, y_{0}, v_{x_{0}}, v_{y_{0}}, v_{z_{0}}\}$ correspond the drifts of the reference frame and the zero point offsets. The rest four parameters $\{\beta,\gamma,\alpha,\lambda\}$ are the expansion parameters of the  PRZ spacetime, as mentioned in the previous section.  Another key point we note is that the S2 star orbit may not be ellipse exactly. However, we presume that the orbit is elliptical, referring to the osculating ellipse with the orbital elements mentioned above.

For the analysis of MCMC simulations together with the above-mentioned parameter space, uniform priors be chosen for all parameters. The ranges of parameters ${\beta,\gamma}$ are set as $[0,2]$ and the ranges of  $\{\alpha,\lambda\}$ are $[-1,1]$. 

It is worth noting here that there exist three various parts of data, which, mentioned in~\ref{Sec:data}, can be applied to the analysis of MCMC simulations. Hence, the probability function $\mathcal{L}$ has three different parts, which are given by 
\begin{eqnarray}
	\log {\cal L} = \log {\cal L}_{\rm AP} + \log {\cal L}_{\rm VR} + \log {\cal L}_{\rm pro}\, ,\label{likelyhood}
\end{eqnarray}
where the first term, $\log {\cal L}_{\rm AP}$, represents the probability of 145 astrometric positional data and is defined by  
\begin{eqnarray}
	\log {\cal L}_{\rm AP} &=& - \frac{1}{2} \sum_{i} \frac{(X_{\rm obs}^i -X_{\rm the}^i)^2}{(\sigma^i_{X,{\rm obs}})^2} \nonumber\\
	&& -\frac{1}{2} \sum_{i} \frac{(Y_{\rm obs}^i -Y_{\rm the}^i)^2}{(\sigma^i_{Y,{\rm obs}})^2}\, ,
\end{eqnarray}
and the second term, $\log {L}_{\rm VR}$, describes the probability of 45 data for the radial velocities and reads as 
\begin{eqnarray}
	\log {L}_{\rm VR} - \frac{1}{2} \sum_{i} \frac{(V_{\rm R, obs}^i - V_{\rm R, the}^i)^2}{(\sigma^i_{V_{\rm R, obs}})^2}\, ,
\end{eqnarray}
while the third term, $\log {\cal L}_{\rm pro}$, represents the probability of the orbital precession and is given by 
\begin{eqnarray}
	\log {\cal L}_{\rm pro} = - \frac{1}{2} \frac{(\Delta \phi_{\rm obs}-\Delta \phi_{\rm the})^2}{\sigma^2_{\Delta \phi, {\rm obs}}}\, ,
\end{eqnarray}
where $X_{\rm obs}^i$, $Y_{\rm obs}^i$, and $V_{\rm R, obs}^i$ respectively refer to the data of the astrometric positions and radial velocities, likewise $X_{\rm the}^i$, $Y_{\rm the}^i$, and $V_{\rm R, the}^i$ refer to the theoretical predictions. Also, $\sigma^i_{x, {\rm obs}^i}$ appearing in the above equation refers to an appropriate statistical uncertainty for the corresponding quantities. Here, we would like to mention the orbital precession $\Delta \phi_{\rm obs}$ of the S2 star given by Eq.~(\ref{OP_12}) and theoretical one $\Delta \phi_{\rm the}$ given by Eq.~(\ref{p.shift2}) involving the above mentioned expansion parameters of the  PRZ spacetime.

\subsection{Results and Discussions}

\begin{figure*}
	\centering
	\includegraphics[width=18.0cm]{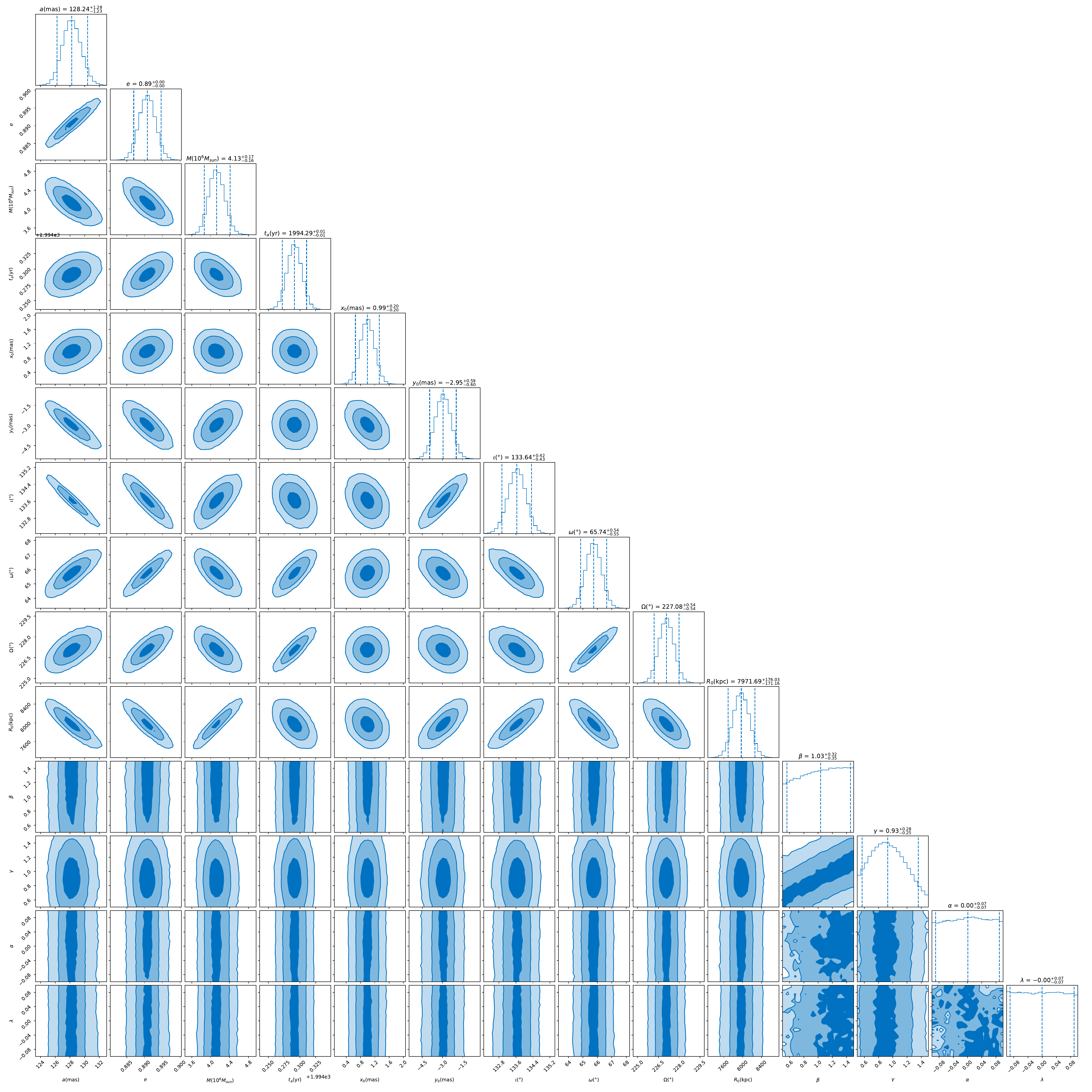}
	\caption{\label{fig:contour} The figure shows the posterior distribution of the S2 star's orbital parameters and the expansion parameters of  PRZ spacetime associated with the uniform priors for these orbital parameters of S2 star.}
\end{figure*}

Following all the above subsections, we analyzed these 14-dimensional parameter spaces by adapting MCMC simulations and showing the posterior distributions of these parameter spaces for the orbital model of S2 star; see Fig.~\ref{fig:contour}. Note that we show the contour plots with {68\%, 90\%, and 95\%} confidence regions. The appropriate constraint values of these parameters are also tabulated in Table.~\ref{table2}. Hence, it is particularly noteworthy because the results shown here in Fig.~\ref{fig:contour} and Table.~\ref{table2} for the first two expansion parameters are well in agreement with the results for PPN. 
\begin{table}
	\caption{\label{table2}
		The constraint the best-fit values of expansion parameters $\{\beta,\gamma,\alpha,\lambda\}$ and the S2 star's orbital model parameters in the  PRZ spacetime are tabulated as stated by the analysis of MCMC simulations. Note that the constraint ranges on the expansion parameters are obtained through the posterior region at {90\%} confidence level.}
	\centering
	%\begin{ruledtabular}
	\begin{tabular}{c c}
		Parameters &\multicolumn{1}{c}{Best-fit values}  \\
		
		\colrule
		$ M\; (10^{6}M_{\odot})$  & {4.127} \\
		$R_{0}$ (kpc)  &{7971} \\
		$a$ (mas)  &{128.24}\\
		$e$   &{0.89077}\\
		$\iota$ ($^{\circ}$)  &{133.64}\\
		$\omega$ ($^{\circ}$) &{65.741}\\
		$\Omega$ ($^{\circ}$)  &{227.07} \\
		$t_{\rm apo}$ (yr)  &{1994.2913}\\
		$x_{0}$ (mas)  &{1.06} \\
		$y_{0}$ (mas)  &{-2.51}\\
		$v_{\rm x_{0}}$ (mas/yr)  & {0.134}\\
		$v_{\rm y_{0}}$ (mas/yr)  & {0.020}\\
		$v_{\rm z_{0}}$ (km/s)  & {13.42}\\
		\colrule
		$\beta$    &  $1.03^{+0.32}_{-0.35}$\\
		$\gamma$    & $0.93^{+0.28}_{-0.25}$\\
		$\alpha$    &  (-0.09,\,0.09)\\ 
		$\lambda$   & (-0.09,\,0.09)\\
		\colrule
	\end{tabular}
	%\end{ruledtabular}
\end{table}
In Fig.~\ref{fig:contour}, we show the observational constraints on the expansion parameters of the  PRZ spacetime by using the data of the S2 star's orbital model.  As a consequence of the analysis, as seen in Fig.~\ref{fig:contour}, we demonstrate the constraint values through the corresponding posterior distribution of the expansion parameters, so the fist two expansion parameters of PRZ spacetime are observationally constrained to be $\beta=1.03^{+0.32}_{-0.35}$ and $\gamma=0.93^{+0.28}_{-0.25}$ at {90\%} confidence level. As stated above, these best-fit constraint values are completely consistent with the values for PPN parameters even though there are a bit stronger in contrast to the ones derived from the observations of the solar system. 
However, we find that the observational constraints of the higher order expansion parameters, $\alpha$, and $\lambda$, through the S2 star are more accurate enough as compared to the one for the SS test observations. This is because $M/r_{+}$ appearing in the third term of Eq.~(\ref{p.shift2}) is of order $\approx10^{-4}$ for the S2 star orbiting around the Sgr A$^{\star}$, thus resulting in having a significant impact on the orbital parameters of the S2 star as compared to the one for SS tests. Hence, we have shown that the observational data for the S2 orbiting around the Sgr A$^{\star}$ do have enough capacity to constrain the last two higher order expansion parameters of the black hole described by the  PRZ spacetime. We must however ensure that whether the constraints that we have obtained are the best-fit constraints on the expansion parameters (i.e., $\alpha$ and $\lambda$). To that one has to consider observational data in the immediate vicinity of the black hole where $M/r_{+}$ should be smaller. Recently, some S-stars (S4711, S62, S4714) orbiting the supermassive black hole in Sgr A* with short orbital periods (7.6 yr$\le$Pb$\le$ 12 yr) were discovered (\cite{Peissker2020ApJ}). These stars have the potential to be utilized for measuring the general relativistic Lense–Thirring (LT) precessions (\cite{Iorio_2020}; \cite{iorio2023possible}). The forthcoming direct measurement of periastron shift of these stars could lead to tighter constraints on our model parameters, even in cases of non-spherically symmetric spacetime. In addition to S-stars, we will also consider the observational data of the selected microquasars, which are well-known as astrophysical quasiperiodic oscillations in the close vicinity of the black holes, where the condition $M/r_{+}\sim 10^{-1}$ is met.  This is what we intend to investigate next in this paper.

%-------------
\section{Limits on the parameters of the PRZ spacetime via the quasiperiodic oscillations\label{QPOs} }

In this section, we are concerned with motion perturbations around stable circular orbits. Such perturbations are quasi-periodic oscillations (QPOs), the frequencies of which have direct observational effects~\citep{McClintock11,Torok05A&A,Barret05,Belloni12,Kotrlova08,Strohmayer01ApJ,Shafee06ApJ,Mustapha20}. We specialize to the case of perturbed circular motion as it represents faithfully the trajectories of in-falling matter in accretion processes.

From now on we consider stable paths in the $\theta=\pi/2$ plane. Epicyclic motion around a stable circular path has two components: one is a radial component in the equatorial plane while another one is a vertical component perpendicular to that plane. Instead of restricting ourselves to a special spherically symmetric metric, we shall for simplicity consider its most general form for further analysis,
\begin{equation}\label{q1}
	ds^2=-g_{tt}(r)dt^2+g_{rr}(r)dr^2+g_{\theta\theta}(r)
	d\Omega^2\, ,
\end{equation}
The details of derivations are given in Refs. ~\citep{Aliev81,Mustapha20} for special static metrics and in Ref.~\citep{Mustapha19} for rotating metrics, upon following the same steps of derivation we arrive at
\begin{eqnarray}
	\label{q2} \nu_r&=&\frac{1}{2\pi}\sqrt{\frac{2 g_{\theta\theta} (g_{tt}')^2-2 g_{tt} g_{tt}' g_{\theta\theta}'-g_{tt} g_{\theta\theta} g_{tt}''}{2 g_{tt} g_{rr} g_{\theta\theta}}+\frac{g_{tt}' g_{\theta\theta}''}{2 g_{rr} g_{\theta\theta}'}}\, ,\\
	\label{q3} \nu_\theta&=&\frac{1}{2\pi}\sqrt{\frac{-g_{tt}'}{g_{\theta\theta}'}}\, ,
\end{eqnarray}
where the prime notation denotes the derivative with respect to $r$. Here $\nu_r$ and $\nu_\theta$ are the frequencies of the perturbed circular motion as detected by an observer at spatial infinity, which are related to the local frequencies, $\Omega_r$ and $\Omega_\theta$, by $\nu_r=\Omega_r/(2\pi u^t)$ and $\nu_\theta=\Omega_\theta/(2\pi u^t)$ where $u^t$ and $u^\varphi$ are the only non-vanishing components of the 4-velocity vector of the in-falling particle. Similar expressions to Eqs.~(\ref{q2}) and~(\ref{q3}) in the presence of a magnetic source were derived in~\citep{Shaymatov22c}.

\begin{figure}[!htb]
	\centering
	\includegraphics[width=0.325\textwidth]{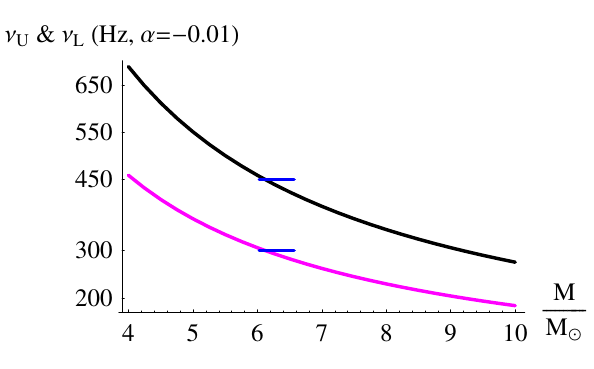}
	\includegraphics[width=0.325\textwidth]{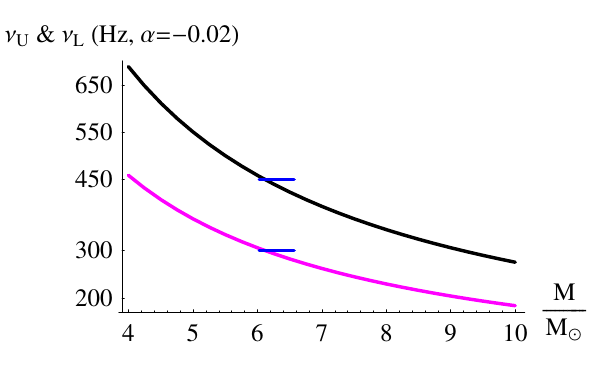}
	\includegraphics[width=0.325\textwidth]{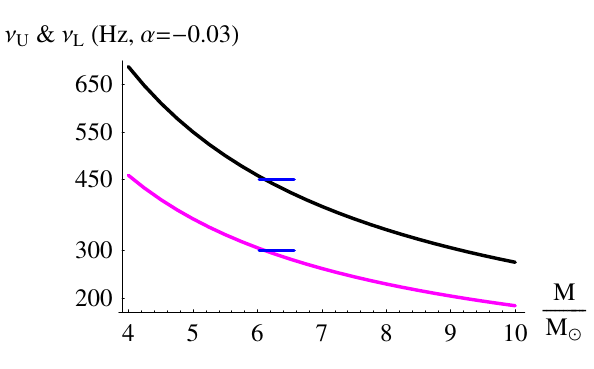}
	
	\includegraphics[width=0.42\textwidth]{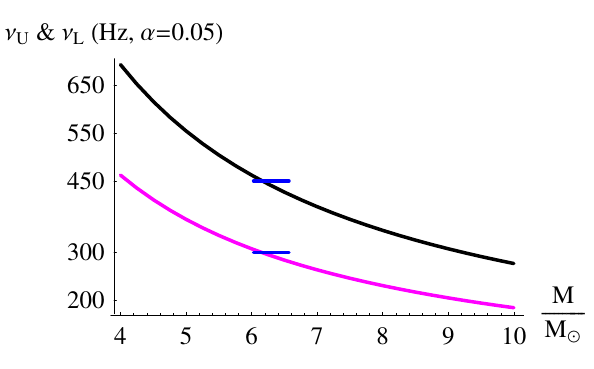}
	\includegraphics[width=0.42\textwidth]{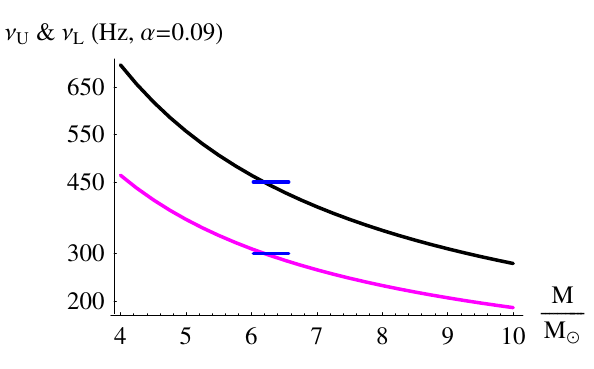}
	\caption{\label{fig:gro}  Curve fit to the data of the GRO J1655-40 galactic microquasar; see Eq.~(\ref{GRO}). The black curves represent $\nu_U=\nu_\theta+\nu_r$, the magenta curves represent $\nu_L=\nu_\theta$, and the blue lines represent the uncertainty on the mass of the GRO J1655-40 microquasar. For theses plots we took $\gamma=1+2.3\times 10^{-5}$, $\beta=1+2.3\times 10^{-4}$, $\lambda=0.06$, and the values of $\alpha$ are shown in the plots.}
\end{figure}

For the case of the spherically symmetric metric we are considering in this work, Eqs.~(\ref{metric}), (\ref{expansion_1}) and~(\ref{expansion_2}), we have
\begin{eqnarray}
	\label{q4} g_{tt}(r)&=&-c^2\Big(1-\frac{2 r_g}{r}+(\beta -\gamma )\frac{2 r_g^2}{r^2}+\alpha \frac{2 r_g^3}{r^3}\Big)\, ,\\ 
	\label{q5} g_{rr}(r)&=&1+\gamma  \frac{2 r_g}{r}+\lambda
	\frac{2 r_g^2}{r^2}\, ,\\
	\label{q6} g_{\theta\theta}(r)&=&r^2\, ,
\end{eqnarray}
where $r_g=GM/c^2$, and $c$ and $G$ are known physical constants taking their values in the SI system to be $299792458$ and $6.673\times 10^{-11}$, respectively. To be consistent with the expansions~(\ref{expansion_1}) and~(\ref{expansion_2}), we only keep the following terms in the expansions of $\nu_r$ and $\nu_\theta$ in terms of the dimensionless variable $y=r/r_g$:
\begin{eqnarray}
	\label{q7}\nu _r&=&\frac{c^3 }{2 \pi  G M y^{3/2}}\Big(1-\frac{2+\gamma }{y}+\frac{3\gamma  (\gamma -4)+3 (\alpha -4)+2 (8 \beta -\lambda )}{2 y^2}\Big)\, ,\\
	\label{q8}\nu _{\theta }&=&\frac{c^3 }{2 \pi  G M y^{3/2}}\Big(1-\frac{\beta -\gamma }{y}-\frac{3 \alpha +(\beta -\gamma )^2}{2 y^2}\Big)\, .
\end{eqnarray}

The two peaks in the power spectra from the GRO J1655-40 galactic microquasar, 
\begin{eqnarray}\label{GRO}
	\text{GRO J1655-40 : }\frac{M}{M_\odot}=6.30\pm 0.27\, ,\;
	\nu_U=450\pm 3 \text{ Hz},\;\nu_L=300\pm 5 \text{ Hz}\, ,
\end{eqnarray}
are at 300 Hz and 450 Hz~\citep{Strohmayer01ApJ}. These two twin values of the QPOs are most certainly due to the phenomenon of resonance which is due to the coupling of non-linear vertical and radial oscillatory motions~\citep{Abramowicz03,Horak06A&A}. The main three models for resonances~\citep{Abramowicz03,Rebusco04,Banerjee22,Deligianni21} are: Parametric resonance, forced resonance and Keplerian resonance. There are other models too~\citep{Banerjee22}. In all three main models, $\nu_U$ and $\nu_L$ are linear combinations of the frequencies $\nu_r$ and $\nu_\theta$ detected by an observer at spatial infinity. Confronting the observed ratio $\nu_U/\nu_L=3/2$ with theory can be done making different assumptions within a resonance model.
\begin{figure}[!htb]
	\centering
	\includegraphics[width=0.325\textwidth]{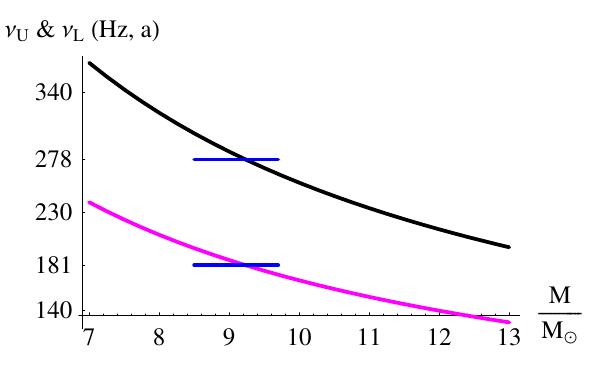}
	\includegraphics[width=0.325\textwidth]{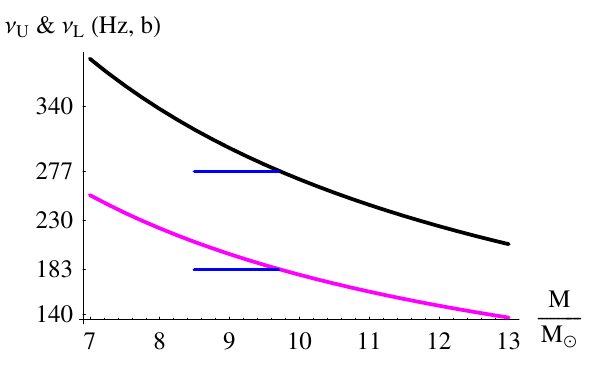}
	\includegraphics[width=0.325\textwidth]{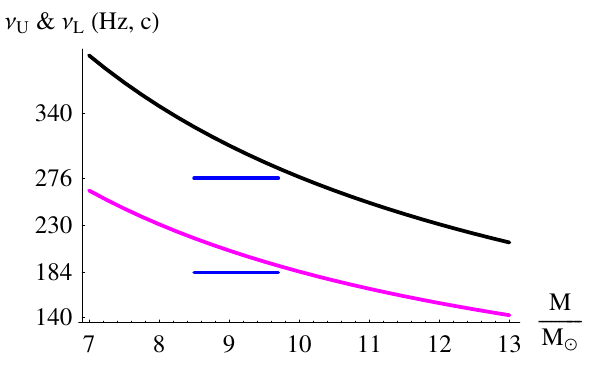}
	\caption{\label{fig:xte}  Curve fit to the data of the XTE J1550-564 galactic microquasar; see Eq.~(\ref{XTE}). The black curves represent $\nu_U=\nu_\theta+\nu_r$, the magenta curves represent $\nu_L=\nu_\theta$, and the blue lines represent the uncertainty on the mass of the XTE J1550-564 microquasar. For these plots we took $\gamma=1+2.3\times 10^{-5}$, $\beta=1+2.3\times 10^{-4}$, $\lambda=0.06$, and $\alpha=0.05$, observing the new constraint, the panels a, b, and c correspond to ($\nu_U=278$ Hz and $\nu_L=181$ Hz), ($\nu_U=277$ Hz, $\nu_L=183$ Hz), and ($\nu_U=276$ Hz, $\nu_L=184$ Hz), respectively.}
\end{figure}

The parameters $\gamma$, $\beta$, $\alpha$, and $\lambda$, as determined in the previous sections, have their values almost identical to those of the Schwarzschild solution. We know from previous studies~\citep{Kolos15qpo} that parametric resonance is not able to justify the two twin peaks at 300 Hz and 450 Hz if the GRO J1655-40 microquasar is modelled by a Schwarzschild black hole unless the effects of a magnetic field surrounding the black hole are taken into consideration. Thus, we opt for forced resonance~\citep{Banerjee22,Deligianni21} with the model
\begin{eqnarray}\label{fr}
	\nu_U=\nu_\theta+\nu_r, \qquad \nu_L=\nu_\theta\, ,
\end{eqnarray}
which provides better results than the other models, i.e., $\nu_U=\nu_\theta$ and $\nu_L=\nu_\theta-\nu_r$. We will be using a graphical presentation, i.e., we plot $\nu_U$ and $\nu_L$ in terms of the mass ratio $M/M_\odot$ for the value of the radial coordinate satisfying $\nu_U/\nu_L=450/300=3/2$ and fixed ($\gamma,\,\beta,\,\alpha$) and we will observe how the curves representing $\nu_U$ and $\nu_L$ cross the mass error band of the GRO J1655-40 microquasar (see for example \citep{Kolos15qpo}). For better results we have selected this microquasar which has the narrowest mass band error $\Delta M=0.54 M_\odot$; see Eq.~(\ref{GRO}). All panels of Fig.~\ref{fig:gro} depict the results clearly. The black curves represent $\nu_U=\nu_\theta+\nu_r$, while the magenta curves represent $\nu_L=\nu_\theta$ given in Eq.~(\ref{fr}) and as well as the blue lines represent the uncertainty on the mass of the GRO J1655-40 microquasar. For theses plots we have taken $\gamma=1+2.3\times 10^{-5}$, $\beta=1+2.3\times 10^{-4}$, $\lambda=0.06$, and the values of $\alpha$ are shown in the plots. From Fig.~\ref{fig:gro}, one can observe that the values of $\alpha$,
\begin{equation}\label{alpha}
	-0.03\leq\alpha\leq 0.09\, ,
\end{equation}
offer a better curve fitting.

Once the constraint Eq.~(\ref{alpha}) is admitted, one can move one step further in constraining rather the parameters of some other microquasars, e.g., the frequencies of the two peaks in the power spectra. The next well-constrained-mass microquasar is XTE J1550-564 with $\Delta M=1.2 M_\odot$, which is given as 
\begin{eqnarray}\label{XTE}
	\text{XTE J1550-564: }\frac{M}{M_\odot}=9.1\pm 0.6\, ,\;
	\nu_U=276\pm 3 \text{ Hz},\;\nu_L=184\pm 5 \text{ Hz}\, .
\end{eqnarray}
Let the mass of the microquasar be in the middle of the mass band, $M=9.1M_\odot$. We assume that the constraints given in Eq.~(\ref{alpha}) hold well. In doing so, we further show in the panel $a$ of Fig.~\ref{fig:xte} that the values
\begin{eqnarray}\label{freq}
	\nu_U=278 \text{ Hz } \text{ and }\; \nu_L=181 \text{ Hz}\, ,	
\end{eqnarray}
offer better curve fitting for the XTE J1550-564 microquasar than the values $\nu_U=276$ Hz and $\nu_L=184$ Hz that are in the middle of the frequency bands, as the panel $c$ of Fig.~\ref{fig:xte} depicts. In the panel $b$, we show how the curve fitting improves progressively upon departing from the frequency-band-middle values and adopting the values $\nu_U=278$ Hz and $\nu_L=181$ Hz.

From the analysis, we can infer that the obtained constraint values through the observations of QPOs in the black hole vicinity also satisfy the best-fit constraint values obtained by applying the observations of the S2 star phenomena around the Sgr A$^{\star}$.

\section{\label{conclusion} Conclusion} 

To test the extended theories of gravity, parametrization does play an important role in “mimicking" various theories of gravity by using expansion of the metric functions in terms of small dimensionless parameters. Therefore, in order to have constraints on the expansion parameters of spacetime from the observation data, it becomes increasingly important to exploit the information obtained from the classical solar system tests, the observations of phenomena of the S2 star located in the star cluster close to the Sgr A$^{\star}$, and microquasars. With this in view, one may be able to obtain information about the spacetime geometry near the horizon. 

In this paper, we expanded functions of the radial coordinate of spherically symmetric  PRZ spacetime and found the higher order expansion parameters $\alpha$ and $\lambda$ that go beyond the first order PPN parameters. We then studied the constraints on the parameters of the spherically symmetric  PRZ spacetime through classical tests of SS effects, the data of the S2 star orbiting the Sgr A$^{\star}$, and the data from GRO J1655-40 and XTE J1550-564 microquasars. We determined the analytical expressions for SS effects, e.g., the perihelion shift, the light deflection, the gravitational time delay, and QPOs' frequencies so as to determine constraints on the higher order expansion parameters of spherically symmetric  PRZ spacetime \citep{Rezzolla14}. We found the constraints on the two expansion parameters $\alpha$ and $\lambda$ that survive only in the vicinity of the horizon by using above mentioned three various observational data and approaches. 

We further considered the impact of the expansion parameters of the  PRZ spacetime on the orbit of S2 star orbiting around Sgr A* at the center of the Milky Way galaxy that can provide excellent tests in probing black hole properties. We also considered the effects of  PRZ spacetime so as to compare with the astrometric and spectroscopic data which is publicly available and involves the astrometric positions, the radial velocities, and the orbital precession for the S2 star considered in this paper. Taking all together we applied the MCMC simulations to probe the possible effects of these expansion parameters on the S2 star orbit and constraints on the parameters of spherically symmetric  PRZ spacetime. We found the best-fit constraint values through the corresponding posterior distribution of the expansion parameters which can be observationally constrained to be $\beta=1.03^{+0.32}_{-0.35}$, $\gamma=0.93^{+0.28}_{-0.25}$ and $\alpha\, ,\lambda=(-0.09\, , 0.09)$  at {90\%} confidence level. However, we found that the observational constraints on $\alpha$, and $\lambda$ are more accurate enough than that of SS tests and have significant impact on the orbit of S2 star. Therefore, it was shown that the observational data for the S2 orbiting around the Sgr A$^{\star}$ is capable to constrain the expansion parameters of the  PRZ spacetime, thus allowing one to obtain the best-fit constraint region on the expansion parameters $\alpha$ and $\lambda$. 

To ensure what we have obtained through the observational data for the S2 we also considered observational data in the close vicinity of the black hole, i.e., $M/r_{+}\sim10^{-1}$ always satisfied, which is comparable with the orbit of S2 star orbiting around Sgr A$^{\star}$. Therefore, we studied the epicyclic motions and derived analytic form of the epicyclic frequencies used to constrain these expansion parameters of  PRZ spacetime by applying the data of the selected microquasars well-known as astrophysical QPOs which are interesting tools for testing and constraining the geometry of metric fields because, as noted in~\citep{Banerjee22}, their frequencies depend solely on the spacetime metric and not on the details of the accretion process. Even the radius of the ISCO where accretion occurs does depend only on the background metric provided the energy-momentum tensor of the fluid is seen as a test matter. From this point of view, QPOs provide deeper insights into the background metric. Thus, observations of QPOs have been used to be very potent tests in probing unknown aspects associated with precise measurements and constraints of the parameters of black holes. Further, we obtained constraints on the higher order expansion parameters of the  PRZ spacetime and constraints on the frequencies of the two peaks in the power spectra of the GRO J1655-40 and XTE J1550-564 microquasars. We showed that the obtained best-fit constraint values through observations of phenomena of the S2 star are well satisfied by the observations of QPOs in the black hole vicinity.    

Finally, taking into consideration all results we can state that these two higher order expansion parameters would be in the range $\alpha\, ,\lambda=(-0.09\, , 0.09)$ and of order $\sim10^{-2}$ as inferred from the observations of SS tests, the S2 star orbiting around the Sgr A$^{\star}$, and QPOs around black hole. As a consequence, our results suggest that the higher order expansion parameters $\alpha$ and $\lambda$ would survive only in the vicinity of black hole horizon in the strong field regime.

\begin{acknowledgments}
	
This work is supported by the National Natural Science Foundation of China under Grants No. 11675143 and No. 11975203, the National Key Research and Development Program of China under Grant No. 2020YFC2201503. B.A. wishes to acknowledge the support from Research F-FA-2021-432 of the Uzbekistan Ministry for Innovative Development. 
	
\end{acknowledgments}

\bibliography{gravreferences,ref}{}

\begin{thebibliography}{}
\expandafter\ifx\csname natexlab\endcsname\relax\def\natexlab#1{#1}\fi
\providecommand{\url}[1]{\href{#1}{#1}}
\providecommand{\dodoi}[1]{doi:~\href{http://doi.org/#1}{\nolinkurl{#1}}}
\providecommand{\doeprint}[1]{\href{http://ascl.net/#1}{\nolinkurl{http://ascl.net/#1}}}
\providecommand{\doarXiv}[1]{\href{https://arxiv.org/abs/#1}{\nolinkurl{https://arxiv.org/abs/#1}}}

\bibitem[{{Abbott} \& et~al. {(Virgo and LIGO Scientific
  Collaborations)}(2016{\natexlab{a}})}]{Abbott16a}
{Abbott}, B.~P., \& et~al. {(Virgo and LIGO Scientific Collaborations)}.
  2016{\natexlab{a}}, Phys. Rev. Lett., 116, 061102,
  \dodoi{10.1103/PhysRevLett.116.061102}

\bibitem[{{Abbott} \& et~al. {(Virgo and LIGO Scientific
  Collaborations)}(2016{\natexlab{b}})}]{Abbott16b}
---. 2016{\natexlab{b}}, Phys. Rev. Lett., 116, 241102,
  \dodoi{10.1103/PhysRevLett.116.241102}

\bibitem[{{Abramowicz} {et~al.}(2003){Abramowicz}, {Karas}, {Kluzniak}, {Lee},
  \& {Rebusco}}]{Abramowicz03}
{Abramowicz}, M.~A., {Karas}, V., {Kluzniak}, W., {Lee}, W.~H., \& {Rebusco},
  P. 2003, Publ. Astron. Soc. Jap., 55, 467, \dodoi{10.1093/pasj/55.2.467}

\bibitem[{{Akiyama} \& et~al. {(Event Horizon Telescope
  Collaboration)}(2019{\natexlab{a}})}]{Akiyama19L1}
{Akiyama}, K., \& et~al. {(Event Horizon Telescope Collaboration)}.
  2019{\natexlab{a}}, Astrophys. J., 875, L1, \dodoi{10.3847/2041-8213/ab0ec7}

\bibitem[{{Akiyama} \& et~al. {(Event Horizon Telescope
  Collaboration)}(2019{\natexlab{b}})}]{Akiyama19L6}
---. 2019{\natexlab{b}}, Astrophys. J., 875, L6,
  \dodoi{10.3847/2041-8213/ab1141}

\bibitem[{{Akiyama} \& et~al. {(Event Horizon Telescope
  Collaboration)}(2022)}]{Akiyama22ApJL}
---. 2022, Astrophys. J. Lett., 930, L12, \dodoi{10.3847/2041-8213/ac6674}

\bibitem[{Aliev {et~al.}(2013)Aliev, Esmer, \& Talazan}]{Aliev12qpo}
Aliev, A.~N., Esmer, G.~D., \& Talazan, P. 2013, Class. Quant. Grav., 30,
  045010, \dodoi{10.1088/0264-9381/30/4/045010}

\bibitem[{{Aliev} \& {Galtsov}(1981)}]{Aliev81}
{Aliev}, A.~N., \& {Galtsov}, D.~V. 1981, Gen. Relativ. Gravit., 13, 899,
  \dodoi{10.1007/BF00756068}

\bibitem[{{Azreg-A{\"\i}nou}(2019)}]{Mustapha19}
{Azreg-A{\"\i}nou}, M. 2019, Int. J. Mod. Phys. D, 28, 1950013,
  \dodoi{10.1142/S0218271819500135}

\bibitem[{{Azreg-A{\"\i}nou} {et~al.}(2020{\natexlab{a}}){Azreg-A{\"\i}nou},
  {Chen}, {Deng}, {Jamil}, {Zhu}, {Wu}, \& {Lim}}]{Azreg-Ainou20qpo}
{Azreg-A{\"\i}nou}, M., {Chen}, Z., {Deng}, B., {et~al.} 2020{\natexlab{a}},
  Phys. Rev. D, 102, 044028, \dodoi{10.1103/PhysRevD.102.044028}

\bibitem[{{Azreg-A{\"\i}nou} {et~al.}(2020{\natexlab{b}}){Azreg-A{\"\i}nou},
  {Chen}, {Deng}, {Jamil}, {Zhu}, {Wu}, \& {Lim}}]{Mustapha20}
---. 2020{\natexlab{b}}, Phys. Rev. D, 102, 044028,
  \dodoi{10.1103/PhysRevD.102.044028}

\bibitem[{{Bambi}(2012)}]{Bambi12a}
{Bambi}, C. 2012, Phys. Rev. D, 85, 043002, \dodoi{10.1103/PhysRevD.85.043002}

\bibitem[{{Bambi} {et~al.}(2016){Bambi}, {Jiang}, \& {Steiner}}]{Bambi16b}
{Bambi}, C., {Jiang}, J., \& {Steiner}, J.~F. 2016, Class. Quant. Grav., 33,
  064001, \dodoi{10.1088/0264-9381/33/6/064001}

\bibitem[{{Banerjee}(2022)}]{Banerjee22}
{Banerjee}, I. 2022, JCAP, 2022, 034, \dodoi{10.1088/1475-7516/2022/08/034}

\bibitem[{{Barret} {et~al.}(2005){Barret}, {Olive}, \& {Miller}}]{Barret05}
{Barret}, D., {Olive}, J.-F., \& {Miller}, M.~C. 2005, Mon. Not. Roy. Astron.
  Soc., 361, 855, \dodoi{10.1111/j.1365-2966.2005.09214.x}

\bibitem[{{Becerra-Vergara} {et~al.}(2020){Becerra-Vergara}, {Arg{\"u}elles},
  {Krut}, {Rueda}, \& {Ruffini}}]{Becerra-Vergara20}
{Becerra-Vergara}, E.~A., {Arg{\"u}elles}, C.~R., {Krut}, A., {Rueda}, J.~A.,
  \& {Ruffini}, R. 2020, Astron. Astrophys., 641, A34,
  \dodoi{10.1051/0004-6361/201935990}

\bibitem[{{Belloni} {et~al.}(2012){Belloni}, {Sanna}, \&
  {M{\'e}ndez}}]{Belloni12}
{Belloni}, T.~M., {Sanna}, A., \& {M{\'e}ndez}, M. 2012, Mon. Not. Roy. Astron.
  Soc., 426, 1701, \dodoi{10.1111/j.1365-2966.2012.21634.x}

\bibitem[{{Bertotti} {et~al.}(2003){Bertotti}, {Iess}, \&
  {Tortora}}]{Bertotti03Nat}
{Bertotti}, B., {Iess}, L., \& {Tortora}, P. 2003, Nature (London), 425, 374,
  \dodoi{10.1038/nature01997}

\bibitem[{{Caldwell} \& {Kamionkowski}(2009)}]{Caldwell09}
{Caldwell}, R., \& {Kamionkowski}, M. 2009, Nature, 458, 587,
  \dodoi{10.1038/458587a}

\bibitem[{{Carloni} {et~al.}(2011){Carloni}, {Grumiller}, \&
  {Preis}}]{Grumiller11}
{Carloni}, S., {Grumiller}, D., \& {Preis}, F. 2011, Phys. Rev. D, 83, 124024,
  \dodoi{10.1103/PhysRevD.83.124024}

\bibitem[{{De Laurentis} {et~al.}(2018){De Laurentis}, {Younsi}, {Porth},
  {Mizuno}, \& {Rezzolla}}]{DeLaurentis18}
{De Laurentis}, M., {Younsi}, Z., {Porth}, O., {Mizuno}, Y., \& {Rezzolla}, L.
  2018, Phys. Rev. D, 97, 104024, \dodoi{10.1103/PhysRevD.97.104024}

\bibitem[{{Deligianni} {et~al.}(2021){Deligianni}, {Kleihaus}, {Kunz},
  {Nedkova}, \& {Yazadjiev}}]{Deligianni21}
{Deligianni}, E., {Kleihaus}, B., {Kunz}, J., {Nedkova}, P., \& {Yazadjiev}, S.
  2021, Phys. Rev. D, 104, 064043, \dodoi{10.1103/PhysRevD.104.064043}

\bibitem[{{Di Valentino} {et~al.}(2021){Di Valentino}, {Mena}, {Pan},
  {Visinelli}, {Yang}, {Melchiorri}, {Mota}, {Riess}, \&
  {Silk}}]{Di_Valentino_2021}
{Di Valentino}, E., {Mena}, O., {Pan}, S., {et~al.} 2021, Class. Quantum
  Gravity, 38, 153001, \dodoi{doi.org/10.1088/1361-6382/ac086d}

\bibitem[{{Do} \& et~al.(2019)}]{Do19}
{Do}, T., \& et~al. 2019, Science, 365, 664, \dodoi{10.1126/science.aav8137}

\bibitem[{Dokuchaev \& Eroshenko(2015)}]{Dokuchaev:2015ghx}
Dokuchaev, V.~I., \& Eroshenko, Y.~N. 2015, Phys. Usp., 58, 772,
  \dodoi{10.3367/UFNe.0185.201508c.0829}

\bibitem[{{Einstein}(1916)}]{Einstein1916}
{Einstein}, A. 1916, Annalen der Physik, 354, 769,
  \dodoi{10.1002/andp.19163540702}

\bibitem[{{Foreman-Mackey} {et~al.}(2013){Foreman-Mackey}, {Hogg}, {Lang}, \&
  {Goodman}}]{emcee}
{Foreman-Mackey}, D., {Hogg}, D.~W., {Lang}, D., \& {Goodman}, J. 2013, Publ.
  Astron. Soc. Pac., 125, 306, \dodoi{10.1086/670067}

\bibitem[{German\`a(2018)}]{Germana18qpo}
German\`a, C. 2018, Phys. Rev. D, 98, 083025,
  \dodoi{10.1103/PhysRevD.98.083025}

\bibitem[{{Ghasemi-Nodehi} {et~al.}(2020){Ghasemi-Nodehi}, {Azreg-A{\"\i}nou},
  {Jusufi}, \& {Jamil}}]{Ghasemi-Nodehi20qpo}
{Ghasemi-Nodehi}, M., {Azreg-A{\"\i}nou}, M., {Jusufi}, K., \& {Jamil}, M.
  2020, Phys. Rev. D, 102, 104032, \dodoi{10.1103/PhysRevD.102.104032}

\bibitem[{Ghez {et~al.}(1998)Ghez, Klein, Morris, \& Becklin}]{Ghez:1998ph}
Ghez, A.~M., Klein, B.~L., Morris, M., \& Becklin, E.~E. 1998, Astrophys. J.,
  509, 678, \dodoi{10.1086/306528}

\bibitem[{{Ghez} {et~al.}(2000){Ghez}, {Morris}, {Becklin}, {Tanner}, \&
  {Kremenek}}]{Ghez00}
{Ghez}, A.~M., {Morris}, M., {Becklin}, E.~E., {Tanner}, A., \& {Kremenek}, T.
  2000, Nature, 407, 349, \dodoi{10.1038/35030032}

\bibitem[{{Ghez} {et~al.}(2005){Ghez}, {Salim}, {Hornstein}, {Tanner}, {Lu},
  {Morris}, {Becklin}, \& {Duch{\^e}ne}}]{Ghez05}
{Ghez}, A.~M., {Salim}, S., {Hornstein}, S.~D., {et~al.} 2005, Astrophys. J.,
  620, 744, \dodoi{10.1086/427175}

\bibitem[{{Gillessen} {et~al.}(2017){Gillessen}, {Plewa}, {Eisenhauer}, {Sari},
  {Waisberg}, {Habibi}, {Pfuhl}, {George}, {Dexter}, {von Fellenberg}, {Ott},
  \& {Genzel}}]{Gillessen17ApJ}
{Gillessen}, S., {Plewa}, P.~M., {Eisenhauer}, F., {et~al.} 2017, Astrophys.
  J., 837, 30, \dodoi{10.3847/1538-4357/aa5c41}

\bibitem[{{Glampedakis} {et~al.}(2017){Glampedakis}, {Pappas}, {Silva}, \&
  {Berti}}]{Glampedakis17}
{Glampedakis}, K., {Pappas}, G., {Silva}, H.~O., \& {Berti}, E. 2017, Phys.
  Rev. D, 96, 064054, \dodoi{10.1103/PhysRevD.96.064054}

\bibitem[{{GRAVITY Collaboration}(2018)}]{GRAVITY1}
{GRAVITY Collaboration}. 2018, Astron. Astrophys., 615, L15,
  \dodoi{10.1051/0004-6361/201833718}

\bibitem[{{GRAVITY Collaboration}(2020)}]{GRAVITY2}
---. 2020, Astron. Astrophys., 636, L5, \dodoi{10.1051/0004-6361/202037813}

\bibitem[{{Grumiller}(2010)}]{Grumiller10}
{Grumiller}, D. 2010, Phys. Rev. Lett., 105, 211303,
  \dodoi{10.1103/PhysRevLett.105.211303}

\bibitem[{{Hellerman} {et~al.}(2001){Hellerman}, {Kaloper}, \&
  {Susskind}}]{Hellerman01}
{Hellerman}, S., {Kaloper}, N., \& {Susskind}, L. 2001, JHEP, 2001, 003,
  \dodoi{10.1088/1126-6708/2001/06/003}

\bibitem[{{Hor{\'a}k} \& {Karas}(2006)}]{Horak06A&A}
{Hor{\'a}k}, J., \& {Karas}, V. 2006, Astron. Astrophys., 451, 377,
  \dodoi{10.1051/0004-6361:20054039}

\bibitem[{{Hulse} \& {Taylor}(1974)}]{Hulse74}
{Hulse}, R.~A., \& {Taylor}, J.~H. 1974, Astrophys. J, 191, L59,
  \dodoi{10.1086/181548}

\bibitem[{{Hulse} \& {Taylor}(1975)}]{Hulse75}
---. 1975, Astrophys. J, 195, L51, \dodoi{10.1086/181708}

\bibitem[{{Iorio}(2015)}]{Iorio15IJMPD}
{Iorio}, L. 2015, Int. J. Mod. Phys. D, 24, 1530015,
  \dodoi{10.1142/S0218271815300153}

\bibitem[{{Iorio}(2019)}]{Iorio2019ApJ}
---. 2019, Astrophys. J., 157, 220, \dodoi{10.3847/1538-3881/ab19bf}

\bibitem[{Iorio(2020)}]{Iorio_2020}
Iorio, L. 2020, Astrophys. J., 904, 186, \dodoi{10.3847/1538-4357/abbfb5}

\bibitem[{Iorio(2023)}]{iorio2023possible}
---. 2023, Is it possible to measure the Lense-Thirring orbital shifts of the
  short-period S-star S4716 orbiting Sgr A$^\ast$?
\newblock \doarXiv{2306.17432}

\bibitem[{{Johannsen} \& {Psaltis}(2011)}]{Johannsen11}
{Johannsen}, T., \& {Psaltis}, D. 2011, Phys. Rev. D, 83, 124015,
  \dodoi{10.1103/PhysRevD.83.124015}

\bibitem[{{Jusufi} {et~al.}(2022{\natexlab{a}}){Jusufi}, {Azreg-A{\"\i}nou},
  {Jamil}, \& {Saridakis}}]{Jusufi22Univ}
{Jusufi}, K., {Azreg-A{\"\i}nou}, M., {Jamil}, M., \& {Saridakis}, E.~N.
  2022{\natexlab{a}}, Universe, 8, 102, \dodoi{10.3390/universe8020102}

\bibitem[{{Jusufi} {et~al.}(2021){Jusufi}, {Azreg-A{\"\i}nou}, {Jamil}, {Wei},
  {Wu}, \& {Wang}}]{Jusufi21PRD}
{Jusufi}, K., {Azreg-A{\"\i}nou}, M., {Jamil}, M., {et~al.} 2021, Phys. Rev. D,
  103, 024013, \dodoi{10.1103/PhysRevD.103.024013}

\bibitem[{{Jusufi} {et~al.}(2022{\natexlab{b}}){Jusufi}, {Kumar},
  {Azreg-A{\"\i}nou}, {Jamil}, {Wu}, \& {Bambi}}]{Jusufi22EPJC}
{Jusufi}, K., {Kumar}, S., {Azreg-A{\"\i}nou}, M., {et~al.} 2022{\natexlab{b}},
  Eur. Phys. J. C, 82, 633, \dodoi{10.1140/epjc/s10052-022-10603-7}

\bibitem[{{Kagramanova} {et~al.}(2006){Kagramanova}, {Kunz}, \&
  {L{\"a}mmerzahl}}]{Kagramanova06}
{Kagramanova}, V., {Kunz}, J., \& {L{\"a}mmerzahl}, C. 2006, Phys. Lett. B,
  634, 465, \dodoi{10.1016/j.physletb.2006.01.069}

\bibitem[{{Kiselev}(2003)}]{Kiselev03}
{Kiselev}, V.~V. 2003, Class. Quantum Grav., 20, 1187

\bibitem[{{Kluzniak} \& {Abramowicz}(2001)}]{Kluzniak01}
{Kluzniak}, W., \& {Abramowicz}, M.~A. 2001, Acta Phys. Pol. B, 32, 3605

\bibitem[{Kolo\v{s} {et~al.}(2015)Kolo\v{s}, Stuchl\'\i{}k, \&
  Tursunov}]{Kolos15qpo}
Kolo\v{s}, M., Stuchl\'\i{}k, Z., \& Tursunov, A. 2015, Class. Quant. Grav.,
  32, 165009, \dodoi{10.1088/0264-9381/32/16/165009}

\bibitem[{{Konoplya} {et~al.}(2016){Konoplya}, {Rezzolla}, \&
  {Zhidenko}}]{Konoplya16}
{Konoplya}, R., {Rezzolla}, L., \& {Zhidenko}, A. 2016, Phys. Rev. D, 93,
  064015, \dodoi{10.1103/PhysRevD.93.064015}

\bibitem[{{Kotrlov{\'a}} {et~al.}(2008){Kotrlov{\'a}}, {Stuchl{\'{\i}}k}, \&
  {T{\"o}r{\"o}k}}]{Kotrlova08}
{Kotrlov{\'a}}, A., {Stuchl{\'{\i}}k}, Z., \& {T{\"o}r{\"o}k}, G. 2008, Class.
  Quantum Grav., 25, 225016, \dodoi{10.1088/0264-9381/25/22/225016}

\bibitem[{{Lacroix}(2018)}]{Lacroix18}
{Lacroix}, T. 2018, Astron. Astrophys., 619, A46,
  \dodoi{10.1051/0004-6361/201832652}

\bibitem[{{McClintock} {et~al.}(2011){McClintock}, {Narayan}, {Davis}, {Gou},
  {Kulkarni}, {Orosz}, {Penna}, {Remillard}, \& {Steiner}}]{McClintock11}
{McClintock}, J.~E., {Narayan}, R., {Davis}, S.~W., {et~al.} 2011, Class.
  Quantum Gravity, 28, 114009, \dodoi{10.1088/0264-9381/28/11/114009}

\bibitem[{{Nampalliwar} {et~al.}(2021){Nampalliwar}, {Kumar}, {Jusufi}, {Wu},
  {Jamil}, \& {Salucci}}]{Nampalliwar21ApJ}
{Nampalliwar}, S., {Kumar}, S., {Jusufi}, K., {et~al.} 2021, Astrophys. J.,
  916, 116, \dodoi{10.3847/1538-4357/ac05cc}

\bibitem[{{Nucita} {et~al.}(2007){Nucita}, {De Paolis}, {Ingrosso}, {Qadir}, \&
  {Zakharov}}]{Nucita07}
{Nucita}, A.~A., {De Paolis}, F., {Ingrosso}, G., {Qadir}, A., \& {Zakharov},
  A.~F. 2007, PASP, 119, 349, \dodoi{10.1086/517934}

\bibitem[{{P{\'a}nis} {et~al.}(2019){P{\'a}nis}, {Kolo{\v{s}}}, \&
  {Stuchl{\'\i}k}}]{Panis19}
{P{\'a}nis}, R., {Kolo{\v{s}}}, M., \& {Stuchl{\'\i}k}, Z. 2019, Eur. Phys. J.
  C, 79, 479, \dodoi{10.1140/epjc/s10052-019-6961-7}

\bibitem[{{Peebles} \& {Ratra}(2003)}]{Peebles03}
{Peebles}, P.~J., \& {Ratra}, B. 2003, Reviews of Modern Physics, 75, 559,
  \dodoi{10.1103/RevModPhys.75.559}

\bibitem[{{Pei{\ss}ker} {et~al.}(2020){Pei{\ss}ker}, {Eckart}, {Zaja{\v{c}}ek},
  {Ali}, \& {Parsa}}]{Peissker2020ApJ}
{Pei{\ss}ker}, F., {Eckart}, A., {Zaja{\v{c}}ek}, M., {Ali}, B., \& {Parsa}, M.
  2020, \apj, 899, 50, \dodoi{10.3847/1538-4357/ab9c1c}

\bibitem[{{Perlick} \& {Tsupko}(2021)}]{Perlick21}
{Perlick}, V., \& {Tsupko}, O.~Y. 2021, arXiv e-prints, arXiv:2105.07101.
\newblock \doarXiv{2105.07101}

\bibitem[{{Persic} {et~al.}(1996){Persic}, {Salucci}, \& {Stel}}]{Persic96}
{Persic}, M., {Salucci}, P., \& {Stel}, F. 1996, Mon. Not. R. Astron. Soc.,
  281, 27, \dodoi{10.1093/mnras/278.1.27}

\bibitem[{{Rayimbaev} {et~al.}(2022){Rayimbaev}, {Majeed}, {Jamil}, {Jusufi},
  \& {Wang}}]{Rayimbaev22qpo}
{Rayimbaev}, J., {Majeed}, B., {Jamil}, M., {Jusufi}, K., \& {Wang}, A. 2022,
  Phys. Dark Universe, 35, 100930, \dodoi{10.1016/j.dark.2021.100930}

\bibitem[{{Rayimbaev} {et~al.}(2021){Rayimbaev}, {Shaymatov}, \&
  {Jamil}}]{Rayimbaev-Shaymatov21a}
{Rayimbaev}, J., {Shaymatov}, S., \& {Jamil}, M. 2021, Eur. Phys. J. C, 81,
  699, \dodoi{10.1140/epjc/s10052-021-09488-9}

\bibitem[{{Rebusco}(2004)}]{Rebusco04}
{Rebusco}, P. 2004, Publ. Astron. Soc. Jap., 56, 553,
  \dodoi{10.1093/pasj/56.3.553}

\bibitem[{{Reid} {et~al.}(2007){Reid}, {Menten}, {Trippe}, {Ott}, \&
  {Genzel}}]{Reid07ApJ}
{Reid}, M.~J., {Menten}, K.~M., {Trippe}, S., {Ott}, T., \& {Genzel}, R. 2007,
  Astrophys. J., 659, 378, \dodoi{10.1086/511744}

\bibitem[{{Rezzolla} {et~al.}(2003){Rezzolla}, {Yoshida}, {Maccarone}, \&
  {Zanotti}}]{Rezzolla_qpo_03a}
{Rezzolla}, L., {Yoshida}, S., {Maccarone}, T.~J., \& {Zanotti}, O. 2003, Mon.
  Not. R. Astron. Soc., 344, L37, \dodoi{10.1046/j.1365-8711.2003.07018.x}

\bibitem[{{Rezzolla} \& {Zhidenko}(2014)}]{Rezzolla14}
{Rezzolla}, L., \& {Zhidenko}, A. 2014, Phys. Rev. D, 90, 084009,
  \dodoi{10.1103/PhysRevD.90.084009}

\bibitem[{{Rubin} {et~al.}(1980){Rubin}, {Ford}, \& {Thonnard}}]{Rubin80}
{Rubin}, V.~C., {Ford}, W.~K., J., \& {Thonnard}, N. 1980, Astrophys. J., 238,
  471, \dodoi{10.1086/158003}

\bibitem[{{Shafee} {et~al.}(2006){Shafee}, {McClintock}, {Narayan}, {Davis},
  {Li}, \& {Remillard}}]{Shafee06ApJ}
{Shafee}, R., {McClintock}, J.~E., {Narayan}, R., {et~al.} 2006, Astrophys. J.
  Lett., 636, L113, \dodoi{10.1086/498938}

\bibitem[{{Shapiro} {et~al.}(2004){Shapiro}, {Davis}, {Lebach}, \&
  {Gregory}}]{Shapiro04PRL}
{Shapiro}, S.~S., {Davis}, J.~L., {Lebach}, D.~E., \& {Gregory}, J.~S. 2004,
  Phys. Rev. Lett., 92, 121101, \dodoi{10.1103/PhysRevLett.92.121101}

\bibitem[{{Shaymatov} {et~al.}(2022){Shaymatov}, {Jamil}, {Jusufi}, \&
  {Bamba}}]{Shaymatov22c}
{Shaymatov}, S., {Jamil}, M., {Jusufi}, K., \& {Bamba}, K. 2022, Eur. Phys. J.
  C, 82, 636, \dodoi{10.1140/epjc/s10052-022-10560-1}

\bibitem[{{Shaymatov} {et~al.}(2020){Shaymatov}, {Vrba}, {Malafarina},
  {Ahmedov}, \& {Stuchl{\'\i}k}}]{Shaymatov20egb}
{Shaymatov}, S., {Vrba}, J., {Malafarina}, D., {Ahmedov}, B., \&
  {Stuchl{\'\i}k}, Z. 2020, Phys. Dark Universe, 30, 100648,
  \dodoi{10.1016/j.dark.2020.100648}

\bibitem[{{Spergel} {et~al.}(2007){Spergel}, {Bean}, \& et~al.
  {(WMAP)}}]{Spergel07}
{Spergel}, D.~N., {Bean}, R., \& et~al. {(WMAP)}. 2007, Astrophys. J. Suppl.,
  170, 377, \dodoi{10.1086/513700}

\bibitem[{{Stella} {et~al.}(1999){Stella}, {Vietri}, \&
  {Morsink}}]{Stella99-qpo}
{Stella}, L., {Vietri}, M., \& {Morsink}, S.~M. 1999, Astrophys. J., 524, L63,
  \dodoi{10.1086/312291}

\bibitem[{{Strohmayer}(2001)}]{Strohmayer01ApJ}
{Strohmayer}, T.~E. 2001, Astrophys. J. Lett., 552, L49, \dodoi{10.1086/320258}

\bibitem[{Stuchlik {et~al.}(2007)Stuchlik, Slany, \& Torok}]{Stuchlik07qpo}
Stuchlik, Z., Slany, P., \& Torok, G. 2007, Astron. Astrophys., 470, 401,
  \dodoi{10.1051/0004-6361:20077051}

\bibitem[{Tarnopolski \& Marchenko(2021)}]{Tarnopolski:2021ula}
Tarnopolski, M., \& Marchenko, V. 2021, Astrophys. J., 911, 20,
  \dodoi{10.3847/1538-4357/abe5b1}

\bibitem[{{Tasheva} \& {Stefanov}(2019)}]{Tasheva18}
{Tasheva}, R.~P., \& {Stefanov}, I.~Z. 2019, in AIP Conf. Proc., Vol. 2075,
  10th Jubilee International Conference of the Balkan Physical Union, 090007,
  \dodoi{10.1063/1.5091221}

\bibitem[{Titarchuk \& Shaposhnikov(2005)}]{Titarchuk05qpo}
Titarchuk, L., \& Shaposhnikov, N. 2005, Astrophys. J., 626, 298,
  \dodoi{10.1086/429986}

\bibitem[{{T{\"o}r{\"o}k} {et~al.}(2005){T{\"o}r{\"o}k}, {Abramowicz},
  {Klu{\'z}niak}, \& {Stuchl{\'\i}k}}]{Torok05A&A}
{T{\"o}r{\"o}k}, G., {Abramowicz}, M.~A., {Klu{\'z}niak}, W., \&
  {Stuchl{\'\i}k}, Z. 2005, Astron. Astrophys., 436, 1,
  \dodoi{10.1051/0004-6361:20047115}

\bibitem[{{Tripathi} {et~al.}(2019){Tripathi}, {Yan}, {Yang}, {Yan}, {Garnham},
  {Yao}, {Li}, {Ding}, {Abdikamalov}, {Ayzenberg}, {Bambi}, {Dauser}, {Garcia},
  {Jiang}, \& {Nampalliwar}}]{Tripathi19}
{Tripathi}, A., {Yan}, J., {Yang}, Y., {et~al.} 2019, arXiv e-prints.
\newblock \doarXiv{1901.03064}

\bibitem[{{Tursunov} {et~al.}(2020){Tursunov}, {Stuchl{\'\i}k}, {Kolo{\v{s}}},
  {Dadhich}, \& {Ahmedov}}]{Tursunov20ApJ}
{Tursunov}, A., {Stuchl{\'\i}k}, Z., {Kolo{\v{s}}}, M., {Dadhich}, N., \&
  {Ahmedov}, B. 2020, Astrophys. J., 895, 14, \dodoi{10.3847/1538-4357/ab8ae9}

\bibitem[{Wald(1984)}]{Wald84}
Wald, R.~M. 1984, General relativity (Chicago: The University of Chicago Press)

\bibitem[{Weinberg(1972)}]{Weinberg72}
Weinberg, S. 1972, Gravitation and Cosmology: Principles and Applications of
  the General Theory of Relativity (New York: John Wiley and Sons)

\bibitem[{{Wetterich}(1988)}]{Wetterich88}
{Wetterich}, C. 1988, Nuclear Physics B, 302, 668,
  \dodoi{10.1016/0550-3213(88)90193-9}

\bibitem[{{Will}(1993)}]{Will93}
{Will}, C.~M. 1993, {Theory and Experiment in Gravitational Physics}, 396

\bibitem[{{Will}(2001)}]{Will01}
---. 2001, Living Rev. Rel., 4, 4, \dodoi{10.12942/lrr-2001-4}

\bibitem[{{Will}(2006)}]{Will06}
---. 2006, Living Rev. Rel., 9, 3, \dodoi{10.12942/lrr-2006-3}

\bibitem[{{Yan} {et~al.}(2022){Yan}, {Wu}, {Liu}, {Zhu}, \& {Wang}}]{Yan22JCAP}
{Yan}, J.-M., {Wu}, Q., {Liu}, C., {Zhu}, T., \& {Wang}, A. 2022, JCAP, 2022,
  008, \dodoi{10.1088/1475-7516/2022/09/008}

\bibitem[{Yang {et~al.}(2021)Yang, Liu, Zhu, Zhao, Wu, Yang, \&
  Jamil}]{Yang:2020bpj}
Yang, S., Liu, C., Zhu, T., {et~al.} 2021, Chin. Phys. C, 45, 015102,
  \dodoi{10.1088/1674-1137/abc066}

\end{thebibliography}
\bibliographystyle{aasjournal}



\end{document}